\documentclass[draft]{agujournal2019}
\usepackage{aas_macros}
\usepackage{url} 
\usepackage{amsmath}
\usepackage{textcomp, gensymb}
\usepackage{amssymb}
\usepackage{longtable}
\usepackage{float}
\usepackage{threeparttable}
\usepackage{booktabs}
\usepackage{makecell}

\usepackage{graphicx}
\usepackage{array}
\usepackage{multirow}

\usepackage{tabularx}
\usepackage{parskip}
\usepackage{inconsolata}
\usepackage{xr} 
\usepackage{soul}
\soulregister\ref7
\soulregister\cite7
\soulregister\citeA7

\draftfalse
\newcommand{\flux}{cm$^{-2}$s$^{-1} $}

\newcommand{\p}{$^+$}
\newcommand{\2}{$_2$}
\newcommand{\3}{$_3$}
\newcommand{\E}[1]{$\times 10^{#1}$}
\journalname{Journal of Geophysical Research: Planets}

\begin{document}

%
%



\title{Fully coupled photochemistry of the deuterated ionosphere of Mars and its effects on escape of H and D}
%
%

\authors{Eryn Cangi\affil{1}, Michael Chaffin\affil{1}, Roger Yelle\affil{2}, Bethan Gregory\affil{1}, Justin Deighan\affil{1}}

\affiliation{1}{Laboratory for Atmospheric and Space Physics, University of Colorado Boulder}
\affiliation{2}{Lunar and Planetary Laboratory, University of Arizona}

\correspondingauthor{Eryn Cangi}{eryn.cangi@colorado.edu}

\begin{keypoints}
    \item We present the first photochemical modeling study of the deuterated ionosphere of Mars.
    \item Non-thermal escape dominates D loss under all solar conditions, and the processes producing hot D are similar to those producing hot H.
    \item The combined D/H fractionation factor is $f=0.04$--0.07, indicating 147--158 m GEL of water loss, still less than geological estimates.
\end{keypoints}

%
%

\begin{abstract}

Although deuterium (D) on Mars has received substantial attention, the deuterated ionosphere remains relatively unstudied. This means that we also know very little about non-thermal D escape from Mars, since it is primarily driven by excess energy imparted to atoms produced in ion-neutral reactions. Most D escape from Mars is expected to be non-thermal, highlighting a gap in our understanding of water loss from Mars. In this work, we set out to fill this knowledge gap. To accomplish our goals, we use an upgraded 1D photochemical model that fully couples ions and neutrals and does not assume photochemical equilibrium. To our knowledge, such a model has not been applied to Mars previously. We model the atmosphere during solar minimum, mean, and maximum, and find that the deuterated ionosphere behaves similarly to the H-bearing ionosphere, but that non-thermal escape on the order of 8000-9000 \flux{} dominates atomic D loss under all solar conditions. The total fractionation factor, $f$, is $f=0.04$--0.07, and integrated water loss is 147--158 m GEL. This is still less than geomorphological estimates. Deuterated ions at Mars are likely difficult to measure with current techniques due to low densities and mass degeneracies with more abundant H ions. Future missions wishing to measure the deuterated ionosphere in situ will need to develop innovative techniques to do so.

\end{abstract}

\section*{Plain Language Summary}

Our knowledge of ions in the martian atmosphere that contain deuterium (D), an isotope of hydrogen (H), is limited, lacking measurements and dedicated computer models. This is a problem because the expectation is that most D that escapes to space does so by gaining extra energy from ion reactions. H and D mostly exist in water on Mars, so identifying how much H and D have escaped this way is important to understanding how Mars lost all the water that it once had, transforming it from a planet that could have supported life into the hostile environment it is today. Here, we present the first one dimensional model of the Mars atmosphere that includes D-bearing ions.  We report the amounts and types of H and D escape and confirm that most D gets its escape energy from ion chemistry. We also identify the specific chemical reactions that are most important, and show how many D-bearing ions we expect to find in the atmosphere that might be detectable by future missions. Finally, we calculate that a layer of water 147--158 m deep has been lost from Mars. This is still less than the amount calculated by geological studies.
%
%

\section{Introduction}

Mars is a natural laboratory to study how atmospheric escape shapes planetary habitability. It is now well established that a significant amount of the Mars atmosphere has been lost to space \cite<e.g.>{Jakosky2018}. This escape is fractionating---the relative escape efficiency is different for members of an isotope pair, such as deuterium (D) and hydrogen (H). Because on Mars, D and H are found primarily in water, D/H fractionation indicates a history of water loss \cite{Owen1988}. Understanding escape fractionation therefore contributes to understanding the long-term loss of the atmosphere and desiccation of the planet. 

Geological studies indicate that Mars has likely lost 500+ meters global equivalent layer (GEL) of water \cite[and references therein]{Lasue2013}, but atmospheric modeling studies typically do not find the same result, instead arriving at a smaller number of 100-250 m GEL \cite{Cangi2020, Alsaeed2019, Kras2002, Kras2000}. A key step in retrieving water loss estimates from atmospheric models is to quantify both thermal and non-thermal escape.

Thermal escape includes hydrodynamic escape, which may have ocurred on early Mars and is not expected to significantly fractionate H and D \cite{Hunten1987}, but does fractionate more massive molecules \cite{Cassata2022, Yoshida2020, Zahnle1990}; and Jeans escape, in which a small fraction of particles escape because they have a thermal velocity in the high-energy tail of the Maxwellian velocity distribution, above planetary escape velocity. Non-thermal escape comprises all other processes that grant extra kinetic energy to atmospheric particles, then variously dubbed ``suprathermal'' or ``hot''; most of these processes involve ion chemistry or interaction with ionsand include sputtering, interactions of the planetary ionosphere with the solar wind, and ion outflow. (Impact-driven escape also falls under the non-thermal category, but doesn't specifically involve ions, and is not expected to be fractionating.) H escape has been well-studied at Mars with atmospheric models, observations from missions, and mixes of the two \cite[and others]{Gu2022, Chaffin2021, Chaufray2021a, Holmes2021, Stone2020, AlMaazmi2019, Kras2019, Mayyasi2018, Rahmati2018, Halekas2017a, Bhattacharyya2017, Zahnle2008, Kras2002}. Thermal escape of D has also been modeled \cite{Cangi2020, Kass1999, Yung1988}, but non-thermal escape of D sourced from planetary ion-neutral reactions has not been modeled, despite indications that non-thermal escape is the dominant loss process of D \cite{Gacesa2012, Kras2010, Kras2002}. \citeA{Kras2010, Kras2002} and \citeA{Kras1998} calculated non-thermal D escape velocities for a few select processes (solar wind charge exchange, electron impact ionization, and photoionization), but their model did not include a deuterated ionosphere, and so missed a portion of the production of hot atoms.

\citeA{Cangi2020} used a 1D photochemical model of Mars' neutral atmosphere to calculate the D/H fractionation factor $f$ as a function of atmospheric temperatures. The model only calculated thermal escape directly; non-thermal escape was approximated by scaling the non-thermal effusion velocities given by \citeA{Kras2002} and multiplying them by the densities of H and D at the exobase. This estimation indicated that $f$ is several orders of magnitude larger when non-thermal escape processes are considered, motivating a more complete calculation of non-thermal escape of H and D. Here, we present this more complete treatment. The key questions about the deuterated martian ionosphere that we address are as follows.

\begin{enumerate}
    \item What are the atmospheric densities of deuterated ions?
    \item What are the dominant production mechanisms of hot H and hot D, and are they analogous or dissimilar?
    \item What is the magnitude of non-thermal escape of D, and is it the dominant type of escape during quiet solar conditions?
    \item Can inclusion of non-thermal escape in the model yield an estimation of water loss similar to the amount calculated in geomorphological studies?
\end{enumerate}

To answer these questions, we have substantially upgraded our existing 1D photochemical model, now named bluejay, of the martian atmosphere. Spanning from the surface to 250 km, it now includes a self-consistent ionosphere with deuterated ions, fully couples the ions and neutrals without assumption of photochemical equilibrium, and has only argon as a fixed background species. To our knowledge, it is the first photochemical model to do all of this in the Mars literature.

Past photochemistry studies have typically used several simplifications which improve computational speed and make it easier to model the chemistry of ions and neutrals, which have very different timescales of chemical reaction. Most recent ion-neutral photochemical models use one or more of three common approaches: (1) a fixed (either wholly or partially) background neutral atmosphere \cite{Fox2021, Fox2017, Fox2015b, Matta2013, MolinaCuberos2002}; (2) placing the lower boundary of the model near the bottom of the ionosphere \cite{Fox2021, Kras2019, Fox2015b, Matta2013, Kras2002}; or (3) the assumption of photochemical equilibrium for chemically short-lived species \cite{Vuitton2019, Banaszkiewicz2000} and/or neglect of ion diffusion \cite{Dobrijevic2016}. By building a model that does not use these approaches, we obtain a more complete understanding of the coupling of the lower to upper atmospheres, which has been recently shown to be key to understanding water transport, destruction, and escape during the Mars dusty season \cite{Villanueva2021, Chaffin2021, Holmes2021, Stone2020, Fedorova2020, Vandaele2019, Aoki2019, Heavens2018}. To accommodate the computational demands of the model, bluejay is built in Julia \cite{Julia}, a relatively new language designed specifically for numerical modeling.

We use our enhanced model to present the first theoretical analysis of D ion chemistry at Mars, which includes an updated quantification of non-thermal escape of D and H, the most critical reactions for production of hot H and D, and the implications for water loss.

\section{Model description} \label{sect:model}

Here we describe changes made to the 1D photochemical model as described by \citeA{Cangi2020}. In addition to the upgrades to physics and chemistry described below, this update incorporates computational improvements, such as extensive encapsulation, vectorization of functions, and performance tuning. The only species that we hold constant in our model is argon and lower atmospheric water (see Section \ref{sect:water}). 

The absolute tolerance is 1 $\times 10^{-12}$, or 1 ppt, and the relative tolerance is 1$\times 10^{-6}$. If the percent change of a value is above relative tolerance, the model continues performing Newton iterations to predict the species densities at the next timestep, and stops once it falls below the relative tolerance. This becomes problematic when the values solved for are near machine precision (for 64-bit floats), as happens for ions in the lower atmosphere. In this scenario, the model relies on the absolute tolerance instead, which represents a level below which we do not demand any numerical accuracy; thus, densities that are below 1 ppt may be less representative of the real atmosphere than densities above.

\subsection{New features}

\subsubsection{Ion reaction network} \label{sect:rates}

Our updated model contains $\sim$600 total ion and neutral reactions. The deuterated reactions are shown in Table \ref{tab:DREACTIONS}, and the non-deuterated reactions are available in Table S2; rate coefficients of H-analogue reactions are generally the same as those used by \citeA{Cangi2020, Vuitton2019, Chaffin2017}.

\textbf{Scope of deuterated reactions:} We define a deuterated analogue reaction as a reaction in which one H atom in one of the reactants has been replaced with D; for example, D + O\2 $\rightarrow$ DO\2 instead of H + O\2 $\rightarrow$ HO\2. We do not consider doubly deuterated reactions or species, e.g., we do not include reactions like DO\2 + D $\rightarrow$ OD + OD nor species like D\2O. The already very low relative abundance of D to H ($\sim1\times 10^{-3}$) means that doubly deuterated species and reactions will have an abundance lower by another factor of $\sim$1000, are thus expected to have a negligible effect on the atmospheric chemistry. Our deuterated reaction network includes all of the reactions shown in Table \ref{tab:DREACTIONS}. The H-bearing analogues of the reactions in Table \ref{tab:DREACTIONS} contribute 99.99997\% of the total column rate of all H-bearing reactions, meaning that most of the atmospheric chemistry of H has a deuterated analogue, so it is unlikely that we have accidentally biased our results against D production. 

\begin{small}
	\begin{longtable}{p{0.5cm}l@{~$\rightarrow$~}lp{0.4cm}p{0.6cm}lp{0.7cm}}
		\caption{Deuterated reactions used in the model. Reactions 1-6b: column rate $\nu$ in \flux{}. Reactions 7-125: rate coefficients in units of cm$^3$ molecule$^{-1}$ s$^{-1}$ for bimolecular reactions and cm$^6$ molecule$^{-1}$ s$^{-1}$ for termolecular reactions. BR = branching ratio; MS = mass scaling.}\label{tab:DREACTIONS} \\
		\toprule
		&  \multicolumn{2}{c}{Reaction} &  BR &                       MS &                                                                                                                                   Rate coefficient &                     Ref \\
		\midrule
		\endfirsthead
		
		\toprule
		& \multicolumn{2}{c}{Reaction}  &  BR &                       MS &                                                                                                                                   Rate or rate coefficient &                     Ref \\
		\midrule
		\endhead
		\midrule
		\multicolumn{7}{r}{{Continued on next page}} \\
		\midrule
		\endfoot
		
		\bottomrule
		\endlastfoot
		\multicolumn{7}{c}{Photodissociation and photoionization} \\
		1 &     $\mathrm{D}$ & $\mathrm{D^+}$           & & & $\nu_{\mathrm{col}}$ = $7.5\times 10^{-12}$ & $\dagger$ \\
		2 &  $\mathrm{DO_2}$ & $\mathrm{OD + O}$        & & & $\nu_{\mathrm{col}}$ = $6.9\times 10^{-8}$   &  $\dagger$ \\
		3a &    $\mathrm{HD}$ & $\mathrm{HD^+}$         & & & $\nu_{\mathrm{col}}$ = $6.2\times 10^{-12}$  &  $\dagger$  \\
		3b &                  & $\mathrm{H + D}$        & & & $\nu_{\mathrm{col}}$ = $3.8\times 10^{-12}$  &  $\dagger$  \\
		3c &                  & $\mathrm{H^+ + D}$      & & & $\nu_{\mathrm{col}}$ = $6.9\times 10^{-13}$  &  $\dagger$  \\
		3d &                  & $\mathrm{D^+ + H}$      & & & $\nu_{\mathrm{col}}$ = $6.9\times 10^{-13}$ &  $\dagger$  \\
		4a & $\mathrm{HDO}$   & $\mathrm{D + OH}$       & & & $\nu_{\mathrm{col}}$ = $4.4\times 10^{-10}$ & C0499 \\
		4b &                  & $\mathrm{H + OD}$       & & & $\nu_{\mathrm{col}}$ = $4.4\times 10^{-10}$ & C0499 \\
		4c &                  & $\mathrm{HD + O(^1D)}$  & & & $\nu_{\mathrm{col}}$ = $5.6\times 10^{-11}$ & C0499 \\
		4d &                  & $\mathrm{HDO^+}$        & & & $\nu_{\mathrm{col}}$ = $3.3\times 10^{-11}$ &  $\dagger$ \\
		4e &                  & $\mathrm{OD^+ + H}$     & & & $\nu_{\mathrm{col}}$ = $6.4\times 10^{-12}$ &  $\dagger$  \\
		4f &                  & $\mathrm{OH^+ + D}$     & & & $\nu_{\mathrm{col}}$ = $6.4\times 10^{-12}$ &  $\dagger$  \\
		4g &                  & $\mathrm{D^+ + OH}$     & & & $\nu_{\mathrm{col}}$ = $2.9\times 10^{-12}$  &  $\dagger$  \\
		4h &                  & $\mathrm{H^+ + OD}$     & & & $\nu_{\mathrm{col}}$ = $2.9\times 10^{-12}$ &  $\dagger$  \\
		4i &                  & $\mathrm{O^+ + HD}$     & & & $\nu_{\mathrm{col}}$ = $3.8\times 10^{-13}$ &  $\dagger$  \\
		4j &                  & $\mathrm{H + D + O}$    & & & $\nu_{\mathrm{col}}$ = 0   &  $\dagger$  \\
		5a & $\mathrm{HDO_2}$ & $\mathrm{OH + OD}$      & & & $\nu_{\mathrm{col}}$ = $1.1\times 10^{-8}$ &  $\dagger$ \\
		5b &                  & $\mathrm{DO_2 + H}$     & & & $\nu_{\mathrm{col}}$ = $3.1\times 10^{-10}$ &  $\dagger$ \\
		5c &                  & $\mathrm{HO_2 + D}$     & & & $\nu_{\mathrm{col}}$ = $3.1\times 10^{-10}$ &  $\dagger$ \\
		5d &                  & $\mathrm{HDO + O(^1D)}$ & & & $\nu_{\mathrm{col}}$ = 0     &  $\dagger$ \\
		6a &    $\mathrm{OD}$ & $\mathrm{O + D}$        & & & $\nu_{\mathrm{col}}$ = $1.1\times 10^{-9}$ & NL84 \\
		6b &                  & $\mathrm{O(^1D) + D}$   & & & $\nu_{\mathrm{col}}$ = $1.6\times 10^{-11}$ & NL84 \\
		\multicolumn{7}{c}{Deuterated neutral-neutral reactions} \\
		7 &          $\mathrm{CO + D}$ &             $\mathrm{DCO}$ &     &                          &              \makecell[l]{See text \\ k$_{\infty}=1\left(T_n\right)^{0.2}$ \\ k$_0=2.00 \times 10^{-35}\left(T_n\right)^{0.2}$} & Est. \\
		8a &         $\mathrm{CO + OD}$ &        $\mathrm{CO_2 + D}$ &     & $\sqrt{ \frac{17}{18} }$ &   \makecell[l]{See text \\ k$_{\infty}=1.63 \times 10^{-6}\left(T_n\right)^{6.1}$ \\ k$_0=4.90 \times 10^{-15}\left(T_n\right)^{0.6}$} & Est. \\
		8b &                            &            $\mathrm{DOCO}$ &     & $\sqrt{ \frac{17}{18} }$ & \makecell[l]{See text \\ k$_{\infty}=6.62 \times 10^{-16}\left(T_n\right)^{1.3}$ \\ k$_0=1.73 \times 10^{-29}\left(T_n\right)^{-1.4}$} & Est. \\
		9 &         $\mathrm{D + H_2}$ &          $\mathrm{HD + H}$ &     &                          &                                                                              $2.73 \times 10^{-17}\left(T_n\right)^{2.0}e^{-2700/T_n}$ & NIST \\
		10a &      $\mathrm{D + H_2O_2}$ &       $\mathrm{H_2O + OD}$ & 0.5 &                          &                                                                                                    $1.16 \times 10^{-11}e^{-2110/T_n}$ &  C10 \\
		10b &                            &        $\mathrm{HDO + OH}$ & 0.5 &                          &                                                                                                    $1.16 \times 10^{-11}e^{-2110/T_n}$ &  C10 \\
		11a &        $\mathrm{D + HO_2}$ &        $\mathrm{DO_2 + H}$ &     &                          &                                                                                                                 $1.00 \times 10^{-10}$ &  Y88 \\
		11b &                            &        $\mathrm{HD + O_2}$ &     &                          &                                                                                                                 $2.45 \times 10^{-12}$ &  Y88 \\
		11c &                            &    $\mathrm{HDO + O(^1D)}$ &     &                          &                                                                                                                 $1.14 \times 10^{-12}$ &  Y88 \\
		11d &                            &         $\mathrm{OH + OD}$ &     &                          &                                                                                                                 $5.11 \times 10^{-11}$ &  Y88 \\
		12 &         $\mathrm{D + O_2}$ &            $\mathrm{DO_2}$ &     &   $\sqrt{ \frac{1}{2} }$ & \makecell[l]{See text \\ k$_{\infty}=2.40 \times 10^{-11}\left(T_n\right)^{0.2}$ \\ k$_0=1.46 \times 10^{-28}\left(T_n\right)^{-1.3}$} & Est. \\
		13 &         $\mathrm{D + O_3}$ &        $\mathrm{OD + O_2}$ &     &                          &                                                                                                     $9.94 \times 10^{-11}e^{-470/T_n}$ & NIST \\
		14 &   $\mathrm{D + OH + CO_2}$ &      $\mathrm{HDO + CO_2}$ &     &   $\sqrt{ \frac{1}{2} }$ &                                                                                          $1.16 \times 10^{-25}\left(T_n\right)^{-2.0}$ & Est. \\
		15 &         $\mathrm{DCO + H}$ &         $\mathrm{CO + HD}$ &     & $\sqrt{ \frac{29}{30} }$ &                                                                                                                 $1.50 \times 10^{-10}$ & Est. \\
		16a &         $\mathrm{DCO + O}$ &         $\mathrm{CO + OD}$ &     & $\sqrt{ \frac{29}{30} }$ &                                                                                                                 $5.00 \times 10^{-11}$ & Est. \\
		16b &                            &        $\mathrm{CO_2 + D}$ &     & $\sqrt{ \frac{29}{30} }$ &                                                                                                                 $5.00 \times 10^{-11}$ & Est. \\
		17a &       $\mathrm{DCO + O_2}$ &       $\mathrm{CO_2 + OD}$ &     & $\sqrt{ \frac{29}{30} }$ &                                                                                                                 $7.60 \times 10^{-13}$ & Est. \\
		17b &                            &       $\mathrm{DO_2 + CO}$ &     & $\sqrt{ \frac{29}{30} }$ &                                                                                                                 $5.20 \times 10^{-12}$ & Est. \\
		18 &        $\mathrm{DCO + OH}$ &        $\mathrm{HDO + CO}$ & 0.5 & $\sqrt{ \frac{29}{30} }$ &                                                                                                                 $1.80 \times 10^{-10}$ & Est. \\
		19 &     $\mathrm{DO_2 + HO_2}$ &     $\mathrm{HDO_2 + O_2}$ &     & $\sqrt{ \frac{33}{34} }$ &                                                                                                      $3.00 \times 10^{-13}e^{460/T_n}$ & Est. \\
		20 &        $\mathrm{DO_2 + N}$ &         $\mathrm{NO + OD}$ &     & $\sqrt{ \frac{33}{34} }$ &                                                                                                                 $2.20 \times 10^{-11}$ & Est. \\
		21 &      $\mathrm{DO_2 + O_3}$ &  $\mathrm{OD + O_2 + O_2}$ &     & $\sqrt{ \frac{33}{34} }$ &                                                                                                     $1.00 \times 10^{-14}e^{-490/T_n}$ & Est. \\
		22 &      $\mathrm{DOCO + O_2}$ &     $\mathrm{DO_2 + CO_2}$ &     & $\sqrt{ \frac{45}{46} }$ &                                                                                                                 $2.09 \times 10^{-12}$ & Est. \\
		23 &       $\mathrm{DOCO + OH}$ &      $\mathrm{CO_2 + HDO}$ &     & $\sqrt{ \frac{45}{46} }$ &                                                                                                                 $1.03 \times 10^{-11}$ & Est. \\
		24 &       $\mathrm{H + D + M}$ &          $\mathrm{HD + M}$ &     &   $\sqrt{ \frac{1}{2} }$ &                                                                                         $6.62 \times 10^{-27}\left(T_n\right)^{-2.27}$ & Est. \\
		25a &        $\mathrm{H + DO_2}$ &        $\mathrm{HD + O_2}$ &     & $\sqrt{ \frac{33}{34} }$ &                                                                                                                 $3.45 \times 10^{-12}$ & Est. \\
		25b &                            &    $\mathrm{HDO + O(^1D)}$ &     & $\sqrt{ \frac{33}{34} }$ &                                                                                                                 $1.60 \times 10^{-12}$ & Est. \\
		25c &                            &        $\mathrm{HO_2 + D}$ &     &                          &                                                                                                     $1.85 \times 10^{-10}e^{-890/T_n}$ &  Y88 \\
		25d &                            &         $\mathrm{OH + OD}$ &     & $\sqrt{ \frac{33}{34} }$ &                                                                                                                 $7.20 \times 10^{-11}$ & Est. \\
		26 &          $\mathrm{H + HD}$ &         $\mathrm{H_2 + D}$ &     &                          &                                                                                                    $1.15 \times 10^{-11}e^{-3041/T_n}$ & NIST \\
		27a &       $\mathrm{H + HDO_2}$ &       $\mathrm{H_2O + OD}$ & 0.5 &                          &                                                                                                    $1.16 \times 10^{-11}e^{-2110/T_n}$ &  C10 \\
		27b &                            &        $\mathrm{HDO + OH}$ & 0.5 &                          &                                                                                                    $1.16 \times 10^{-11}e^{-2110/T_n}$ &  C10 \\
		28 &   $\mathrm{H + OD + CO_2}$ &      $\mathrm{HDO + CO_2}$ &     & $\sqrt{ \frac{17}{18} }$ &                                                                                          $1.16 \times 10^{-25}\left(T_n\right)^{-2.0}$ & Est. \\
		29 &         $\mathrm{HCO + D}$ &         $\mathrm{CO + HD}$ &     &   $\sqrt{ \frac{1}{2} }$ &                                                                                                                 $1.50 \times 10^{-10}$ & Est. \\
		30 &        $\mathrm{HCO + OD}$ &        $\mathrm{HDO + CO}$ & 0.5 & $\sqrt{ \frac{29}{30} }$ &                                                                                                                 $1.80 \times 10^{-10}$ & Est. \\
		31a &          $\mathrm{HD + O}$ &          $\mathrm{OD + H}$ &     &                          &                                                                                                    $1.68 \times 10^{-12}e^{-4400/T_n}$ & NIST \\
		31b &                            &          $\mathrm{OH + D}$ &     &                          &                                                                                                    $4.40 \times 10^{-12}e^{-4390/T_n}$ & NIST \\
		32 & $\mathrm{HO_2 + DO_2 + M}$ & $\mathrm{HDO_2 + O_2 + M}$ &     & $\sqrt{ \frac{33}{34} }$ &                                                                                                      $4.20 \times 10^{-33}e^{920/T_n}$ & Est. \\
		33 &       $\mathrm{HOCO + OD}$ &      $\mathrm{CO_2 + HDO}$ &     & $\sqrt{ \frac{17}{18} }$ &                                                                                                                 $1.03 \times 10^{-11}$ & Est. \\
		34 &           $\mathrm{O + D}$ &              $\mathrm{OD}$ &     &   $\sqrt{ \frac{1}{2} }$ &                                                                                         $8.65 \times 10^{-18}\left(T_n\right)^{-0.38}$ & Est. \\
		35 &        $\mathrm{O + DO_2}$ &        $\mathrm{OD + O_2}$ &     & $\sqrt{ \frac{33}{34} }$ &                                                                                                      $3.00 \times 10^{-11}e^{200/T_n}$ & Est. \\
		36a &       $\mathrm{O + HDO_2}$ &       $\mathrm{OD + HO_2}$ & 0.5 & $\sqrt{ \frac{34}{35} }$ &                                                                                                    $1.40 \times 10^{-12}e^{-2000/T_n}$ & Est. \\
		36b &                            &       $\mathrm{OH + DO_2}$ & 0.5 & $\sqrt{ \frac{34}{35} }$ &                                                                                                    $1.40 \times 10^{-12}e^{-2000/T_n}$ & Est. \\
		37 &          $\mathrm{O + OD}$ &         $\mathrm{O_2 + D}$ &     & $\sqrt{ \frac{17}{18} }$ &                                                                                                      $1.80 \times 10^{-11}e^{180/T_n}$ & Est. \\
		38a &     $\mathrm{O(^1D) + HD}$ &          $\mathrm{D + OH}$ &     &                          &                                                                                                                 $4.92 \times 10^{-11}$ &  Y88 \\
		38b &                            &          $\mathrm{H + OD}$ &     &                          &                                                                                                                 $4.92 \times 10^{-11}$ &  Y88 \\
		39 &    $\mathrm{O(^1D) + HDO}$ &         $\mathrm{OD + OH}$ &     & $\sqrt{ \frac{18}{19} }$ &                                                                                                       $1.63 \times 10^{-10}e^{60/T_n}$ & Est. \\
		40 &          $\mathrm{OD + H}$ &          $\mathrm{OH + D}$ &     &                          &                                                                              $4.58 \times 10^{-9}\left(T_n\right)^{-0.63}e^{-717/T_n}$ &  Y88 \\
		41 &        $\mathrm{OD + H_2}$ &         $\mathrm{HDO + H}$ &     &                          &                                                                                                    $2.80 \times 10^{-12}e^{-1800/T_n}$ &  Y88 \\
		42 &     $\mathrm{OD + H_2O_2}$ &      $\mathrm{HDO + HO_2}$ &     & $\sqrt{ \frac{17}{18} }$ &                                                                                                     $2.90 \times 10^{-12}e^{-160/T_n}$ & Est. \\
		43 &       $\mathrm{OD + HO_2}$ &       $\mathrm{HDO + O_2}$ &     & $\sqrt{ \frac{17}{18} }$ &                                                                                                      $4.80 \times 10^{-11}e^{250/T_n}$ & Est. \\
		44 &        $\mathrm{OD + O_3}$ &      $\mathrm{DO_2 + O_2}$ &     & $\sqrt{ \frac{17}{18} }$ &                                                                                                     $1.70 \times 10^{-12}e^{-940/T_n}$ & Est. \\
		45a &         $\mathrm{OD + OH}$ &         $\mathrm{HDO + O}$ &     & $\sqrt{ \frac{17}{18} }$ &                                                                                                                 $1.80 \times 10^{-12}$ & Est. \\
		45b &                            &           $\mathrm{HDO_2}$ &     & $\sqrt{ \frac{17}{18} }$ &                       \makecell[l]{See text \\ k$_{\infty}=2.60 \times 10^{-11}$ \\ k$_0=6.60 \times 10^{-29}\left(T_n\right)^{-0.8}$} & Est. \\
		46 &          $\mathrm{OH + D}$ &          $\mathrm{OD + H}$ &     &                          &                                                                                          $3.30 \times 10^{-9}\left(T_n\right)^{-0.63}$ &  Y88 \\
		47 &       $\mathrm{OH + DO_2}$ &       $\mathrm{HDO + O_2}$ &     & $\sqrt{ \frac{33}{34} }$ &                                                                                                      $4.80 \times 10^{-11}e^{250/T_n}$ & Est. \\
		48a &         $\mathrm{OH + HD}$ &        $\mathrm{H_2O + D}$ &     &                          &                                                                                                    $4.20 \times 10^{-13}e^{-1800/T_n}$ &  Y88 \\
		48b &                            &         $\mathrm{HDO + H}$ &     &                          &                                                                                                    $5.00 \times 10^{-12}e^{-2130/T_n}$ &  S11 \\
		49a &      $\mathrm{OH + HDO_2}$ &     $\mathrm{H_2O + DO_2}$ & 0.5 & $\sqrt{ \frac{34}{35} }$ &                                                                                                     $2.90 \times 10^{-12}e^{-160/T_n}$ & Est. \\
		49b &                            &      $\mathrm{HDO + HO_2}$ & 0.5 & $\sqrt{ \frac{34}{35} }$ &                                                                                                     $2.90 \times 10^{-12}e^{-160/T_n}$ & Est. \\
		\multicolumn{7}{c}{Deuterated ion-neutral reactions} \\ 
		  50 &     $\mathrm{ArD^+ + CO}$ &     $\mathrm{DCO^+ + Ar}$ &      &                          &                                       $1.25 \times 10^{-9}$ &       A03 \\
		51 &   $\mathrm{ArD^+ + CO_2}$ &   $\mathrm{DCO_2^+ + Ar}$ &      &                          &                                       $1.10 \times 10^{-9}$ &       A03 \\
		52a &    $\mathrm{ArD^+ + H_2}$ &     $\mathrm{ArH^+ + HD}$ &      &                          &                                      $4.50 \times 10^{-10}$ &       A03 \\
		52b &               $\mathrm{}$ &    $\mathrm{H_2D^+ + Ar}$ &      &                          &                                      $8.80 \times 10^{-10}$ &       A03 \\
		53 &    $\mathrm{ArD^+ + N_2}$ &    $\mathrm{N_2D^+ + Ar}$ &      &                          &                                      $6.00 \times 10^{-10}$ &       A03 \\
		54 &     $\mathrm{ArH^+ + HD}$ &    $\mathrm{H_2D^+ + Ar}$ &      &                          &                                      $8.60 \times 10^{-10}$ &       A03 \\
		55a &      $\mathrm{Ar^+ + HD}$ &      $\mathrm{ArD^+ + H}$ &      &                          &                                      $3.84 \times 10^{-10}$ &       A03 \\
		55b &               $\mathrm{}$ &      $\mathrm{ArH^+ + D}$ &      &                          &                                      $3.68 \times 10^{-10}$ &       A03 \\
		55c &               $\mathrm{}$ &      $\mathrm{HD^+ + Ar}$ &      &                          &                                      $4.80 \times 10^{-11}$ &       A03 \\
		56a &     $\mathrm{CO_2^+ + D}$ &      $\mathrm{DCO^+ + O}$ &      &                          &                                      $6.38 \times 10^{-11}$ &       A03 \\
		56b &               $\mathrm{}$ &     $\mathrm{D^+ + CO_2}$ &      &                          &                                      $2.02 \times 10^{-11}$ &       A03 \\
		57a &    $\mathrm{CO_2^+ + HD}$ &    $\mathrm{DCO_2^+ + H}$ &  0.5 &   $\sqrt{ \frac{2}{3} }$ &                                      $2.35 \times 10^{-10}$ &      Est. \\
		57b &               $\mathrm{}$ &    $\mathrm{HCO_2^+ + D}$ &  0.5 &   $\sqrt{ \frac{2}{3} }$ &                                      $2.35 \times 10^{-10}$ &      Est. \\
		58a &   $\mathrm{CO_2^+ + HDO}$ &   $\mathrm{DCO_2^+ + OH}$ &  0.5 & $\sqrt{ \frac{18}{19} }$ &                                      $3.00 \times 10^{-10}$ &      Est. \\
		58b &               $\mathrm{}$ &   $\mathrm{HCO_2^+ + OD}$ &  0.5 & $\sqrt{ \frac{18}{19} }$ &                                      $3.00 \times 10^{-10}$ &      Est. \\
		58c &               $\mathrm{}$ &   $\mathrm{HDO^+ + CO_2}$ &      & $\sqrt{ \frac{18}{19} }$ &                                       $1.80 \times 10^{-9}$ &      Est. \\
		59 &       $\mathrm{CO^+ + D}$ &       $\mathrm{D^+ + CO}$ &      &                          &                                      $9.00 \times 10^{-11}$ &       A03 \\
		60a &      $\mathrm{CO^+ + HD}$ &      $\mathrm{DCO^+ + H}$ & 0.25 &   $\sqrt{ \frac{2}{3} }$ &                                      $7.50 \times 10^{-10}$ &      Est. \\
		60b &               $\mathrm{}$ &      $\mathrm{DOC^+ + H}$ & 0.25 &   $\sqrt{ \frac{2}{3} }$ &                                      $7.50 \times 10^{-10}$ &      Est. \\
		60c &               $\mathrm{}$ &      $\mathrm{HCO^+ + D}$ & 0.25 &   $\sqrt{ \frac{2}{3} }$ &                                      $7.50 \times 10^{-10}$ &      Est. \\
		60d &               $\mathrm{}$ &      $\mathrm{HOC^+ + D}$ & 0.25 &   $\sqrt{ \frac{2}{3} }$ &                                      $7.50 \times 10^{-10}$ &      Est. \\
		61a &     $\mathrm{CO^+ + HDO}$ &     $\mathrm{DCO^+ + OH}$ &  0.5 & $\sqrt{ \frac{18}{19} }$ &                                      $8.40 \times 10^{-10}$ &      Est. \\
		61b &               $\mathrm{}$ &     $\mathrm{HCO^+ + OD}$ &  0.5 & $\sqrt{ \frac{18}{19} }$ &                                      $8.40 \times 10^{-10}$ &      Est. \\
		61c &               $\mathrm{}$ &     $\mathrm{HDO^+ + CO}$ &      & $\sqrt{ \frac{18}{19} }$ &                                       $1.56 \times 10^{-9}$ &      Est. \\
		62 &       $\mathrm{C^+ + HD}$ &       $\mathrm{CH^+ + D}$ & 0.17 &                          &                                      $1.20 \times 10^{-16}$ &       A03 \\
		63a &      $\mathrm{C^+ + HDO}$ &      $\mathrm{DCO^+ + H}$ &  0.5 & $\sqrt{ \frac{18}{19} }$ &                $7.80 \times 10^{-9}\left(T_i\right)^{-0.5}$ &      Est. \\
		63b &               $\mathrm{}$ &      $\mathrm{DOC^+ + H}$ &  0.5 & $\sqrt{ \frac{18}{19} }$ &                                       $1.08 \times 10^{-9}$ &      Est. \\
		63c &               $\mathrm{}$ &      $\mathrm{HCO^+ + D}$ &  0.5 & $\sqrt{ \frac{18}{19} }$ &                $7.80 \times 10^{-9}\left(T_i\right)^{-0.5}$ &      Est. \\
		63d &               $\mathrm{}$ &      $\mathrm{HDO^+ + C}$ &      &                          &                                      $2.34 \times 10^{-10}$ &      Est. \\
		63e &               $\mathrm{}$ &      $\mathrm{HOC^+ + D}$ &  0.5 & $\sqrt{ \frac{18}{19} }$ &                                       $1.08 \times 10^{-9}$ &      Est. \\
		64 &   $\mathrm{DCO_2^+ + CO}$ &   $\mathrm{DCO^+ + CO_2}$ &      & $\sqrt{ \frac{45}{46} }$ &                                      $7.80 \times 10^{-10}$ &      Est. \\
		65a &  $\mathrm{DCO_2^+ + e^-}$ &         $\mathrm{CO + O}$ & 0.68 &                          &               $4.62 \times 10^{-5}\left(T_i\right)^{-0.64}$ &       G05 \\
		65b &               $\mathrm{}$ &        $\mathrm{CO + OD}$ & 0.27 &                          &               $4.62 \times 10^{-5}\left(T_i\right)^{-0.64}$ &       G05 \\
		65c &               $\mathrm{}$ &       $\mathrm{CO_2 + D}$ & 0.05 &                          &               $4.62 \times 10^{-5}\left(T_i\right)^{-0.64}$ &       G05 \\
		66 & $\mathrm{DCO_2^+ + H_2O}$ & $\mathrm{H_2DO^+ + CO_2}$ &      & $\sqrt{ \frac{45}{46} }$ &                                       $2.65 \times 10^{-9}$ &      Est. \\
		67 &    $\mathrm{DCO_2^+ + O}$ &    $\mathrm{DCO^+ + O_2}$ &      & $\sqrt{ \frac{45}{46} }$ &                                      $5.80 \times 10^{-10}$ &      Est. \\
		68a &    $\mathrm{DCO^+ + e^-}$ &         $\mathrm{CO + D}$ & 0.92 &                          &                $9.02 \times 10^{-5}\left(T_i\right)^{-1.1}$ &        GK \\
		68b &               $\mathrm{}$ &         $\mathrm{OD + C}$ & 0.07 &                          &                $9.02 \times 10^{-5}\left(T_i\right)^{-1.1}$ &        GK \\
		69 &      $\mathrm{DCO^+ + H}$ &      $\mathrm{HCO^+ + D}$ &      &                          &                                      $1.50 \times 10^{-11}$ &       A03 \\
		70 &   $\mathrm{DCO^+ + H_2O}$ &   $\mathrm{H_2DO^+ + CO}$ &      & $\sqrt{ \frac{29}{30} }$ &                                       $2.60 \times 10^{-9}$ &      Est. \\
		71 &     $\mathrm{DOC^+ + CO}$ &     $\mathrm{DCO^+ + CO}$ &      & $\sqrt{ \frac{29}{30} }$ &                                      $6.00 \times 10^{-10}$ &      Est. \\
		72 &    $\mathrm{DOC^+ + e^-}$ &         $\mathrm{OD + C}$ &      & $\sqrt{ \frac{29}{30} }$ &                 $1.19 \times 10^{-8}\left(T_i\right)^{1.2}$ &      Est. \\
		73a &    $\mathrm{DOC^+ + H_2}$ &    $\mathrm{H_2D^+ + CO}$ & 0.57 &                          &                                      $6.20 \times 10^{-10}$ &       A03 \\
		73b &               $\mathrm{}$ &     $\mathrm{HCO^+ + HD}$ & 0.43 &                          &                                      $6.20 \times 10^{-10}$ &       A03 \\
		74a &     $\mathrm{D^+ + CO_2}$ &     $\mathrm{CO_2^+ + D}$ &      &                          &                                       $3.50 \times 10^{-9}$ &       A03 \\
		74b &               $\mathrm{}$ &      $\mathrm{DCO^+ + O}$ &      &                          &                                       $2.60 \times 10^{-9}$ &       A03 \\
		75 &        $\mathrm{D^+ + H}$ &        $\mathrm{D + H^+}$ & 0.87 &                          &                $6.50 \times 10^{-11}\left(T_i\right)^{0.5}$ &       Y89 \\
		76 &      $\mathrm{D^+ + H_2}$ &       $\mathrm{H^+ + HD}$ &      &                          &                                       $2.20 \times 10^{-9}$ &       A03 \\
		77a &     $\mathrm{D^+ + H_2O}$ &     $\mathrm{H_2O^+ + D}$ &      &                          &                                       $5.20 \times 10^{-9}$ &       A03 \\
		77b &               $\mathrm{}$ &      $\mathrm{HDO^+ + H}$ &  0.5 &   $\sqrt{ \frac{1}{2} }$ &                                       $8.20 \times 10^{-9}$ &      Est. \\
		78 &       $\mathrm{D^+ + NO}$ &       $\mathrm{NO^+ + D}$ &      &                          &                                       $1.80 \times 10^{-9}$ &       A03 \\
		79 &        $\mathrm{D^+ + O}$ &        $\mathrm{D + O^+}$ &      &                          &                                      $2.80 \times 10^{-10}$ &       A03 \\
		80 &      $\mathrm{D^+ + O_2}$ &      $\mathrm{O_2^+ + D}$ &      &                          &                                       $1.60 \times 10^{-9}$ &       A03 \\
		81a &  $\mathrm{H_2DO^+ + e^-}$ &        $\mathrm{H_2 + O}$ &  0.5 & $\sqrt{ \frac{19}{20} }$ &                $9.68 \times 10^{-8}\left(T_i\right)^{-0.5}$ &      Est. \\
		81b &               $\mathrm{}$ &       $\mathrm{H_2O + D}$ &  0.5 & $\sqrt{ \frac{19}{20} }$ &                $1.86 \times 10^{-6}\left(T_i\right)^{-0.5}$ &      Est. \\
		81c &               $\mathrm{}$ &         $\mathrm{HD + O}$ &  0.5 & $\sqrt{ \frac{19}{20} }$ &                $9.68 \times 10^{-8}\left(T_i\right)^{-0.5}$ &      Est. \\
		81d &               $\mathrm{}$ &        $\mathrm{HDO + H}$ &  0.5 & $\sqrt{ \frac{19}{20} }$ &                $1.86 \times 10^{-6}\left(T_i\right)^{-0.5}$ &      Est. \\
		81e &               $\mathrm{}$ &         $\mathrm{OD + H}$ &  0.5 & $\sqrt{ \frac{19}{20} }$ &                $4.47 \times 10^{-6}\left(T_i\right)^{-0.5}$ &      Est. \\
		81f &               $\mathrm{}$ &       $\mathrm{OD + H_2}$ &  0.5 & $\sqrt{ \frac{19}{20} }$ &                $1.04 \times 10^{-6}\left(T_i\right)^{-0.5}$ &      Est. \\
		81g &               $\mathrm{}$ &         $\mathrm{OH + D}$ &  0.5 & $\sqrt{ \frac{19}{20} }$ &                $4.47 \times 10^{-6}\left(T_i\right)^{-0.5}$ &      Est. \\
		81h &               $\mathrm{}$ &        $\mathrm{OH + HD}$ &  0.5 & $\sqrt{ \frac{19}{20} }$ &                $1.04 \times 10^{-6}\left(T_i\right)^{-0.5}$ &      Est. \\
		82a &    $\mathrm{H_2D^+ + CO}$ &    $\mathrm{DCO^+ + H_2}$ & 0.33 &                          &                                       $1.60 \times 10^{-9}$ &       A03 \\
		82b &               $\mathrm{}$ &     $\mathrm{HCO^+ + HD}$ & 0.67 &                          &                                       $1.60 \times 10^{-9}$ &       A03 \\
		83a &   $\mathrm{H_2D^+ + e^-}$ &          $\mathrm{H + H}$ & 0.73 &                          &                                       $6.00 \times 10^{-8}$ & L96 \\
		83b &               $\mathrm{}$ &        $\mathrm{H_2 + D}$ & 0.07 &                          &                                       $6.00 \times 10^{-8}$ & L96 \\
		83c &               $\mathrm{}$ &         $\mathrm{HD + H}$ &  0.2 &                          &                                       $6.00 \times 10^{-8}$ & L96 \\
		84 &   $\mathrm{H_2D^+ + H_2}$ &     $\mathrm{H_3^+ + HD}$ &      &                          &                                      $5.30 \times 10^{-10}$ &       A03 \\
		85a &    $\mathrm{H_2O^+ + HD}$ &    $\mathrm{H_2DO^+ + H}$ &  0.5 &   $\sqrt{ \frac{2}{3} }$ &                                      $3.80 \times 10^{-10}$ &      Est. \\
		85b &               $\mathrm{}$ &     $\mathrm{H_3O^+ + D}$ &  0.5 &   $\sqrt{ \frac{2}{3} }$ &                                      $3.80 \times 10^{-10}$ &      Est. \\
		86 &  $\mathrm{HCO_2^+ + HDO}$ & $\mathrm{H_2DO^+ + CO_2}$ &      & $\sqrt{ \frac{18}{19} }$ &                                       $2.65 \times 10^{-9}$ &      Est. \\
		87 &      $\mathrm{HCO^+ + D}$ &      $\mathrm{DCO^+ + H}$ &      &                          &                                      $4.25 \times 10^{-11}$ &       A03 \\
		88 &    $\mathrm{HCO^+ + HDO}$ &   $\mathrm{H_2DO^+ + CO}$ &      & $\sqrt{ \frac{18}{19} }$ &                                       $2.60 \times 10^{-9}$ &      Est. \\
		89a &     $\mathrm{HDO^+ + CO}$ &     $\mathrm{DCO^+ + OH}$ &  0.5 & $\sqrt{ \frac{18}{19} }$ &                                      $2.12 \times 10^{-10}$ &      Est. \\
		89b &               $\mathrm{}$ &     $\mathrm{HCO^+ + OD}$ &  0.5 & $\sqrt{ \frac{18}{19} }$ &                                      $2.12 \times 10^{-10}$ &      Est. \\
		90a &    $\mathrm{HDO^+ + e^-}$ &         $\mathrm{HD + O}$ &  0.1 &                          &                                       $1.50 \times 10^{-7}$ &  J99 \\
		90b &               $\mathrm{}$ &          $\mathrm{O + D}$ & 0.59 &                          &                                       $1.50 \times 10^{-7}$ &  J99 \\
		90c &               $\mathrm{}$ &         $\mathrm{OD + H}$ & 0.21 &                          &                                       $1.50 \times 10^{-7}$ &  J99 \\
		90d &               $\mathrm{}$ &         $\mathrm{OH + D}$ &  0.1 &                          &                                       $1.50 \times 10^{-7}$ &  J99 \\
		91a &    $\mathrm{HDO^+ + H_2}$ &    $\mathrm{H_2DO^+ + H}$ &  0.5 & $\sqrt{ \frac{18}{19} }$ &                                      $3.80 \times 10^{-10}$ &      Est. \\
		91b &               $\mathrm{}$ &     $\mathrm{H_3O^+ + D}$ &  0.5 & $\sqrt{ \frac{18}{19} }$ &                                      $3.80 \times 10^{-10}$ &      Est. \\
		92a &      $\mathrm{HDO^+ + N}$ &      $\mathrm{HNO^+ + D}$ &  0.5 & $\sqrt{ \frac{18}{19} }$ &                                      $5.60 \times 10^{-11}$ &      Est. \\
		92b &               $\mathrm{}$ &      $\mathrm{NO^+ + HD}$ &      & $\sqrt{ \frac{18}{19} }$ &                                      $2.80 \times 10^{-11}$ &      Est. \\
		93 &     $\mathrm{HDO^+ + NO}$ &     $\mathrm{NO^+ + HDO}$ &      & $\sqrt{ \frac{18}{19} }$ &                                      $4.60 \times 10^{-10}$ &      Est. \\
		94 &      $\mathrm{HDO^+ + O}$ &     $\mathrm{O_2^+ + HD}$ &      & $\sqrt{ \frac{18}{19} }$ &                                      $4.00 \times 10^{-11}$ &      Est. \\
		95 &    $\mathrm{HDO^+ + O_2}$ &    $\mathrm{O_2^+ + HDO}$ &      & $\sqrt{ \frac{18}{19} }$ &                                      $3.30 \times 10^{-10}$ &      Est. \\
		96a &      $\mathrm{HD^+ + Ar}$ &      $\mathrm{ArD^+ + H}$ & 0.45 &   $\sqrt{ \frac{2}{3} }$ &                                       $2.10 \times 10^{-9}$ &       A03 \\
		96b &               $\mathrm{}$ &      $\mathrm{ArH^+ + D}$ & 0.55 &   $\sqrt{ \frac{2}{3} }$ &                                       $2.10 \times 10^{-9}$ &      Est. \\
		97a &      $\mathrm{HD^+ + CO}$ &      $\mathrm{DCO^+ + H}$ &  0.5 &   $\sqrt{ \frac{2}{3} }$ &                                       $1.45 \times 10^{-9}$ &      Est. \\
		97b &               $\mathrm{}$ &      $\mathrm{HCO^+ + D}$ &  0.5 &   $\sqrt{ \frac{2}{3} }$ &                                       $1.45 \times 10^{-9}$ &      Est. \\
		98a &    $\mathrm{HD^+ + CO_2}$ &    $\mathrm{DCO_2^+ + H}$ &  0.5 &   $\sqrt{ \frac{2}{3} }$ &                                       $1.17 \times 10^{-9}$ &      Est. \\
		98b &               $\mathrm{}$ &    $\mathrm{HCO_2^+ + D}$ &  0.5 &   $\sqrt{ \frac{2}{3} }$ &                                       $1.17 \times 10^{-9}$ &      Est. \\
		99 &     $\mathrm{HD^+ + e^-}$ &          $\mathrm{H + D}$ &      &                          & $1.93 \times 10^{-6}\left(T_i\right)^{-0.853}e^{-43.3/T_i}$ &      KIDA \\
		100 &      $\mathrm{HD^+ + HD}$ &     $\mathrm{H_2D^+ + D}$ &      &                          &                                      $8.42 \times 10^{-10}$ &       A03 \\
		101a &     $\mathrm{HD^+ + N_2}$ &     $\mathrm{N_2D^+ + H}$ &  0.5 &   $\sqrt{ \frac{2}{3} }$ &                                       $1.00 \times 10^{-9}$ &      Est. \\
		101b &               $\mathrm{}$ &     $\mathrm{N_2H^+ + D}$ &  0.5 &   $\sqrt{ \frac{2}{3} }$ &                                       $1.00 \times 10^{-9}$ &      Est. \\
		102a &       $\mathrm{HD^+ + O}$ &       $\mathrm{OD^+ + H}$ &  0.5 &   $\sqrt{ \frac{2}{3} }$ &                                      $7.50 \times 10^{-10}$ &      Est. \\
		102b &               $\mathrm{}$ &       $\mathrm{OH^+ + D}$ &  0.5 &   $\sqrt{ \frac{2}{3} }$ &                                      $7.50 \times 10^{-10}$ &      Est. \\
		103 &     $\mathrm{HD^+ + O_2}$ &     $\mathrm{HO_2^+ + D}$ &  0.5 &   $\sqrt{ \frac{2}{3} }$ &                                      $9.60 \times 10^{-10}$ &      Est. \\
		104 &       $\mathrm{H^+ + HD}$ &      $\mathrm{D^+ + H_2}$ &      &                          &                                      $1.10 \times 10^{-10}$ &       A03 \\
		105a &      $\mathrm{H^+ + HDO}$ &     $\mathrm{H_2O^+ + D}$ &  0.5 & $\sqrt{ \frac{18}{19} }$ &                                       $8.20 \times 10^{-9}$ &      Est. \\
		105b &               $\mathrm{}$ &      $\mathrm{HDO^+ + H}$ &  0.5 & $\sqrt{ \frac{18}{19} }$ &                                       $8.20 \times 10^{-9}$ &      Est. \\
		106 &    $\mathrm{N_2D^+ + CO}$ &    $\mathrm{DCO^+ + N_2}$ &      & $\sqrt{ \frac{29}{30} }$ &                                      $8.80 \times 10^{-10}$ &      Est. \\
		107 &   $\mathrm{N_2D^+ + e^-}$ &        $\mathrm{N_2 + D}$ &      & $\sqrt{ \frac{29}{30} }$ &               $6.60 \times 10^{-7}\left(T_i\right)^{-0.51}$ &      Est. \\
		108 &     $\mathrm{N_2D^+ + H}$ &     $\mathrm{N_2H^+ + D}$ &      &                          &                                      $2.50 \times 10^{-11}$ &       A03 \\
		109 &     $\mathrm{N_2D^+ + O}$ &     $\mathrm{OD^+ + N_2}$ &      & $\sqrt{ \frac{29}{30} }$ &                                      $1.40 \times 10^{-10}$ &      Est. \\
		110 &     $\mathrm{N_2H^+ + D}$ &     $\mathrm{N_2D^+ + H}$ &      &                          &                                      $8.00 \times 10^{-11}$ &       A03 \\
		111 &      $\mathrm{N_2^+ + D}$ &      $\mathrm{D^+ + N_2}$ &      &                          &                                      $1.20 \times 10^{-10}$ &       A03 \\
		112a &     $\mathrm{N_2^+ + HD}$ &     $\mathrm{N_2D^+ + H}$ & 0.51 &                          &                                       $1.34 \times 10^{-9}$ &       A03 \\
		112b &               $\mathrm{}$ &     $\mathrm{N_2H^+ + D}$ & 0.49 &                          &                                       $1.34 \times 10^{-9}$ &       A03 \\
		113a &    $\mathrm{N_2^+ + HDO}$ &    $\mathrm{HDO^+ + N_2}$ &      & $\sqrt{ \frac{18}{19} }$ &                                       $1.90 \times 10^{-9}$ &      Est. \\
		113b &               $\mathrm{}$ &    $\mathrm{N_2D^+ + OH}$ &  0.5 & $\sqrt{ \frac{18}{19} }$ &                                      $5.04 \times 10^{-10}$ &      Est. \\
		113c &               $\mathrm{}$ &    $\mathrm{N_2H^+ + OD}$ &  0.5 & $\sqrt{ \frac{18}{19} }$ &                                      $5.04 \times 10^{-10}$ &      Est. \\
		114 &       $\mathrm{N^+ + HD}$ &       $\mathrm{NH^+ + D}$ & 0.25 &                          &                                      $3.10 \times 10^{-10}$ &       A03 \\
		115 &      $\mathrm{OD^+ + CO}$ &      $\mathrm{DCO^+ + O}$ &      & $\sqrt{ \frac{17}{18} }$ &                                      $8.40 \times 10^{-10}$ &      Est. \\
		116 &    $\mathrm{OD^+ + CO_2}$ &    $\mathrm{DCO_2^+ + O}$ &      & $\sqrt{ \frac{17}{18} }$ &                                       $1.35 \times 10^{-9}$ &      Est. \\
		117 &     $\mathrm{OD^+ + e^-}$ &          $\mathrm{O + D}$ &      & $\sqrt{ \frac{17}{18} }$ &                $6.50 \times 10^{-7}\left(T_i\right)^{-0.5}$ &      Est. \\
		118a &     $\mathrm{OD^+ + H_2}$ &     $\mathrm{H_2O^+ + D}$ &  0.5 & $\sqrt{ \frac{17}{18} }$ &                                      $9.70 \times 10^{-10}$ &      Est. \\
		118b &               $\mathrm{}$ &      $\mathrm{HDO^+ + H}$ &  0.5 & $\sqrt{ \frac{17}{18} }$ &                                      $9.70 \times 10^{-10}$ &      Est. \\
		119 &       $\mathrm{OD^+ + N}$ &       $\mathrm{NO^+ + D}$ &      & $\sqrt{ \frac{17}{18} }$ &                                      $8.90 \times 10^{-10}$ &      Est. \\
		120 &     $\mathrm{OD^+ + N_2}$ &     $\mathrm{N_2D^+ + O}$ &      & $\sqrt{ \frac{17}{18} }$ &                                      $2.40 \times 10^{-10}$ &      Est. \\
		121 &       $\mathrm{OD^+ + O}$ &      $\mathrm{O_2^+ + D}$ &      & $\sqrt{ \frac{17}{18} }$ &                                      $7.10 \times 10^{-10}$ &      Est. \\
		122 &     $\mathrm{OD^+ + O_2}$ &     $\mathrm{O_2^+ + OD}$ &      & $\sqrt{ \frac{17}{18} }$ &                                      $3.80 \times 10^{-10}$ &      Est. \\
		123a &      $\mathrm{OH^+ + HD}$ &     $\mathrm{H_2O^+ + D}$ &      &   $\sqrt{ \frac{2}{3} }$ &                                      $9.70 \times 10^{-10}$ &      Est. \\
		123b &               $\mathrm{}$ &      $\mathrm{HDO^+ + H}$ &      &   $\sqrt{ \frac{2}{3} }$ &                                      $9.70 \times 10^{-10}$ &      Est. \\
		124 &        $\mathrm{O^+ + D}$ &        $\mathrm{D^+ + O}$ &      &   $\sqrt{ \frac{1}{2} }$ &                                      $6.40 \times 10^{-10}$ &      Est. \\
		125a &       $\mathrm{O^+ + HD}$ &       $\mathrm{OD^+ + H}$ & 0.46 &                          &                                       $1.25 \times 10^{-9}$ &       A03 \\
		125b &               $\mathrm{}$ &       $\mathrm{OH^+ + D}$ & 0.54 &                          &                                       $1.25 \times 10^{-9}$ &       A03 \\
		126 &      $\mathrm{O^+ + HDO}$ &      $\mathrm{HDO^+ + O}$ &      & $\sqrt{ \frac{18}{19} }$ &                                       $2.60 \times 10^{-9}$ &      Est. \\
		\hline

		\multicolumn{7}{l}{
			$\dagger$: Cross section assumed same as H-analogue.
			A03: \citeA{Anicich2003}. 
			
		} \\
	 	\multicolumn{7}{l}{
			C0499: \citeA{Cheng2004, Cheng1999}.
			C10: \citeA{Cazaux2010}.			
		} \\
		\multicolumn{7}{l}{ 
			G05: \citeA{Geppert2005}.
			K09: \citeA{Korolov2009}.
			GK: Rate from K09, branching ratio from G05.
		} \\
		\multicolumn{7}{l}{    
			J99: \citeA{Jensen1999}.
			KIDA: \citeA{KIDA}.
			L96: \citeA{Larsson1996}.
		} \\ 
		\multicolumn{7}{l}{  
			NIST: \citeA{NIST}. 
			NL84: \citeA{Nee1984}. 
			S11: \citeA{Sander2011}.
		} \\
		\multicolumn{7}{l}{
			Y88: \citeA{Yung1988}. 
			Y89: \citeA{Yung1989}.
			Est: Estimated with mass scaling (see text).
		}
	\end{longtable}
\end{small}

\textbf{Photodissociation and photoionization}: Photodissociation and ionization of deuterated species is calculated using the solar spectrum (see Section \ref{sect:insolation}), so the entry in the table under `Rate or rate coefficient' represents the integrated column rate. The `Ref' column refers to the source of the cross sections used. For photoionization cross sections of the H-analogue reactions, see \citeA[and references therein]{Vuitton2019}.

\textbf{Neutral and ion bimolecular and termolecular reactions}: The rate coefficient used for a given reaction is the product of the `BR', `MS' and `Rate coefficient' columns (empty fields are taken to be 1). `BR', or branching ratio, accounts for the fact that deuteration of a reaction can create two or more branches with differing products where only one branch would exist for the H-analogue reaction. `MS', or mass scaling, is a scaling factor equal to the square root of the mass ratio, $\sqrt{m_1/m_2}$, where $m_2$ is the mass of the deuterated species and $m_1$ the H-bearing species. This factor is applied to reactions for which we were not able to find a measurement in the literature to account for replacement of one reactant H atom with one D atom; a similar approach was used previously by \citeA{Kras2002} for reactions of neutral HD with dominant ions and minor H-bearing ions.

Most reactions in these tables proceed using the listed rate coefficients. A few exceptions apply; the categorization Types and formulae mentioned below are the same as used by \citeA{Vuitton2019}. A more complete description of the formulae used can be found in their Appendix B.

\textit{Reaction 7}: Similar to its analogue CO + H $\rightarrow$ HCO, this is a Type 4 (pressure dependent association) reaction. The Troe parameter for this reaction is 0, so we use the form:

\begin{equation}
     k = k_R + \frac{(M k_0 k_{\infty})}{M k_0 + k_{\infty}}
\end{equation}

Where $k_R$ is 0 in this case and M is the background atmospheric density.

\textit{Reactions 8a, 8b, 12, and 45b}: These are Type 6 (CO + OD $\rightarrow$ CO\2 + D) and Type 5 (CO + OH $\rightarrow$ DOCO, D + O\2$\rightarrow$ DO\2, OD + OH $\rightarrow$ HDO\2) pressure dependent bimolecular reactions, with the formulae originally given by \citeA{Burkholder2019, Sander2011}. We use the same forms here, but multiplied by our mass scaling factor.

Ion reactions which produce a lone D or H atom have the potential to cause the produced atom to be ``hot'', that is, gaining enough excess energy from the reaction that they can escape. We describe this in more detail in Section \ref{sect:ntesc}.

\subsubsection{Ambipolar diffusion} The model employs ambipolar diffusion for all ions, using the Langevin-Gioumousis-Stevenson equation \cite{Bauer1973}:

\begin{gather}
    D_{ai} = \frac{k(T_i+T_e)}{m_i\sum \nu_{ij}} \\
    \nu_{ij} = 2\pi \left(\frac{\alpha_j e^2}{\mu_{ij}}\right)^{1/2}n_j
\end{gather}

Where $D_{ai}$ is the ambipolar diffusion coefficient for ion $i$, $\nu_{ij}$ is the collision frequency of ion $i$ with neutral $j$, $\alpha_j$ is the polarizability, $e$ is the fundamental charge, and $n$ is the neutral density. Polarizability values for neutrals are collected from \citeA{NIST}. Where polarizability was not available either in data or models for a deuterated species we include, we assumed the same value as the H-bearing analogue.

\subsubsection{Partially fixed water profile} \label{sect:water}

We assume a constant abundance of water in the lower atmosphere, which approximates the average water available due to seasonal cycles of polar cap sublimation and transport. The mixing ratio is 1.3 \E{-4} up to the hygropause; in our model (which does not include cloud microphysics), this is the point at which the water mixing ratio begins to follow the saturation vapor pressure curve. We take the hygropause to be 40 km, between 25 km by \citeA{Kras2002} and its enhanced altitude of 50-80 km during dust storms \cite{Heavens2018} and roughly in the middle of the cloud-forming region \cite{Daerden2022,Neary2020}.  At 72 km, a minimum of saturation is reached; above that level, the abundance of water is a free variable. This allows a more holistic understanding of water and water ion chemistry in the upper atmosphere, which has been shown to be an important tracer of seasonal H escape \cite{Stone2020}. The total amount of water in the atmosphere is 10.5 precipitable $\mu$m, in accordance with observations \cite{Smith2004, Jakosky1982}.

\subsubsection{Non-thermal escape} \label{sect:ntesc}

Although there are many non-thermal escape mechanisms, in this work, we focus on photochemical loss, i.e. the contribution to escape from chemistry and photochemistry. We neglect processes involving the solar wind such as sputtering, ion pickup, and charge exchange with the solar wind. Processes which depend upon the solar wind will primarily occur above the bow shock (which is far above our top boundary), where the solar wind can interact with the corona before being mostly deflected around the planet \cite{Halekas2017b}. By focusing on planetary ionospheric reactions, we capture the non-thermal escape of H and D sourced from the atmosphere below the exobase.

We calculate the non-thermal escape of hot atoms created via ion-neutral chemistry as the product of the probability of escape and the volume production rate of hot atoms using the procedure described by \citeA{Gregory2023a}, using the collisional cross sections for D and H on O (lacking data on collision with CO\2) by \citeA{Zhang2009}. We have evaluated all ion-neutral reactions that produce H, D, H\2, or HD in the model for their exothermicity ($\Delta H_{\mathrm{f,products}}^o - \Delta H_{\mathrm{f,reactants}}^o$, following \citeA{Fox2015a}) and only use those where the excess energy, equal to the difference in enthalpies of products and reactants, is positive and exceeds the energy required for H or D to escape to space. In reality, varying amounts of the excess heat produced can contribute to electronic and/or rovibrational excitations of the more massive second product molecule. A detailed analysis of these energy branches is beyond the scope of this work, in which we provide a first look at these reactions, so we neglect branching of energy into the non-H or D product, and assume that all excess energy ends up in the atomic H or D (see the Supporting Information). We use the escape probability curve calculated by \citeA{Gregory2023a} for an H atom of excess energy 5 eV; this is a reasonable approximation of the actual mean excess energy in our model, which is 3.6 eV. At this time, escape probabilities for larger atoms and molecules have not been calculated, so we also apply the escape probability to D, H\2, and HD. We expect this may introduce no more than a few percent uncertainty in our calculation of the escape of these species. A more rigorous analysis of energy branching ratios and their effect on escape, as well as an update with probability curves for D, H\2, and HD, is a worthy subject for future work.

The resulting volume escape rate can be integrated to obtain an escape flux for the top boundary of the model. Although our focus is escape of atomic H and D, some loss does occur via loss of the molecular form, so we also include non-thermal escape of H\2 and HD. In these cases, we assume that $\mathrm{\sigma_{H_2}}$ is the same as for D due to the similar masses, and that $\mathrm{\sigma_{HD}}$ is larger than H\2 by the same amount that D is larger than H. Although this is not an ideal assumption if one is interested in highly accurate non-thermal H\2 and HD escape, it is acceptable for our purposes, as that escape is very small compared to atomic escape of H and D.

\subsection{Model inputs} 

\begin{figure}[h]
\includegraphics[width=\linewidth]{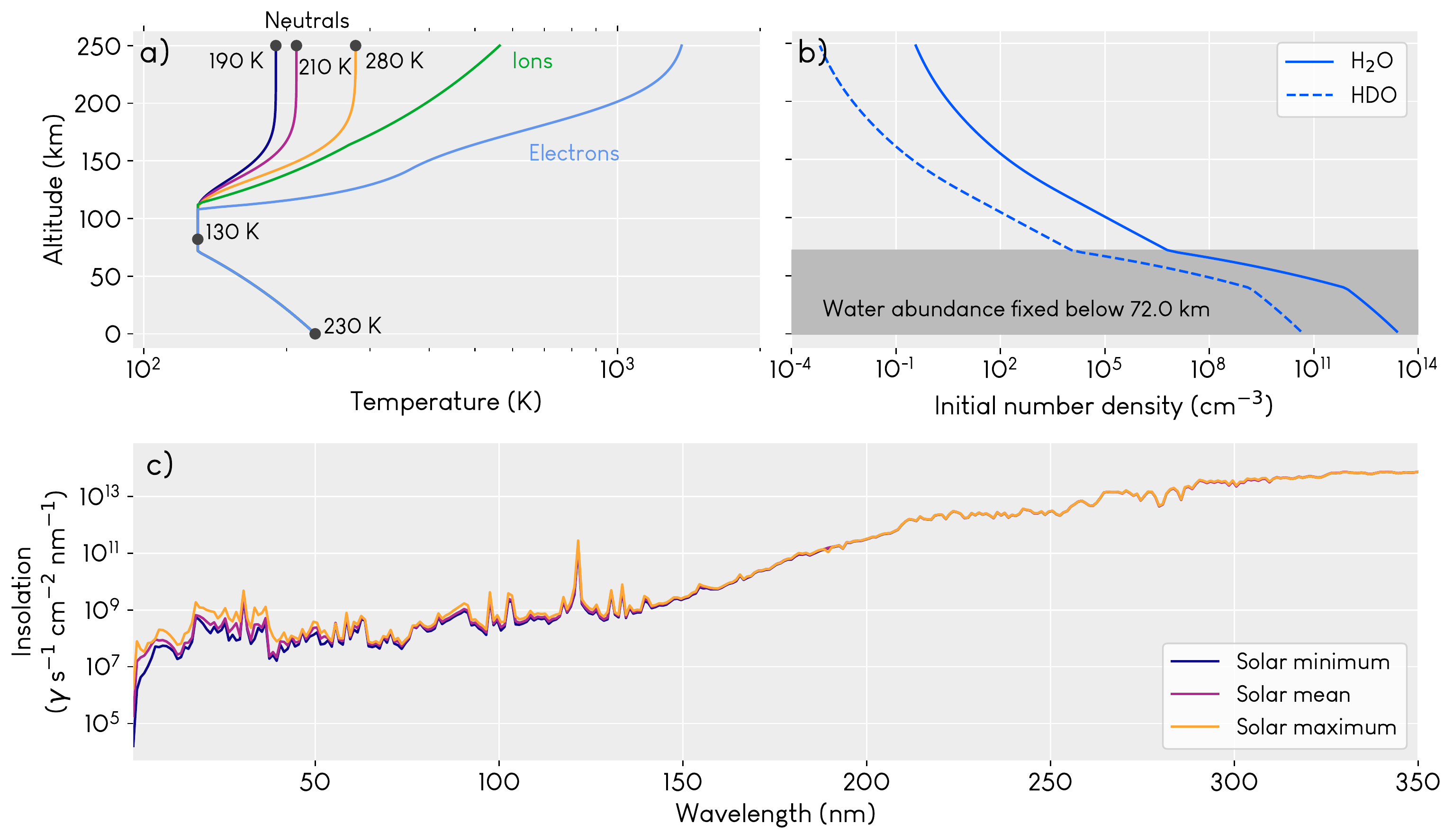}
\caption{Main model inputs. a) Temperature profiles, with separate neutral exobase temperatures for each solar condition. Ion and electron temperatures are fits to data from MAVEN/STATIC as reported by \citeA{Hanley2022} and MAVEN/LPW as reported by \citeA{Ergun2015}. b) Initial water profile. Above 72 km, water densities evolve according to the chemistry and transport. c) Insolation profiles from 0-350 nm for solar minimum, mean, and maximum. The full input spectrum goes out to 2400 nm, but the insolation there is relatively flat, with no variation due to solar cycle.}
\label{inputs}
\end{figure}

Because the importance of non-thermal escape is expected to vary with solar activity, we have constructed three sets of inputs representing solar minimum, mean, and maximum conditions. The only properties which we vary between these cases are the neutral exobase temperatures and the incoming solar flux. Figure \ref{inputs} shows these inputs in the navy, purple, and yellow colors. The inputs represent a dayside mean atmosphere (solar zenith angle [SZA]=60\degree).

\subsubsection{Atmospheric temperature profiles}

Standard neutral temperatures were obtained from the Mars Climate Database \cite{MCD} by several layers of averaging, in order of first to last: by longitude, local time (9, 12, and 3 pm local times, night excluded), latitude (weighted by encompassed surface area), and L$_s$. Over the solar cycle, the only significant change is to the exobase temperature, so we hold the surface and mesospheric temperature constant at 230 K and 130 K respectively.

In order to support modeling of ion chemistry, we use a piecewise fit to the new ion temperature profiles obtained at SZA=60$\degree$ with the STATIC instrument by \citeA{Hanley2022}. These new data have overturned long-standing assumptions that the neutrals, ions, and electrons thermalize to the same temperature around 125 km \cite{SchunkandNagy2009}, and thus represent a significant update in Mars photochemistry. We also include a fit to the electron profile from the Mars Atmosphere and Volatile EvolutioN mission (MAVEN) Langmuir Probes and Waves (LPW) instrument \cite{Ergun2015}. Because it is difficult to associate ion temperatures with contemporary neutral temperatures due to the averaging required for the neutral profiles, and because the data are limited in time, we do not change the ion or electron profiles for the different solar cycle scenarios. Past work indicates relatively minor temperature differences due to solar cycle \cite{Kim1998}; further analysis of MAVEN STATIC data provides an opportunity to confirm or deny this. Any variation with solar cycle is likely to have a minor impact on our results.

\subsubsection{Insolation} \label{sect:insolation}

Incoming solar photons are key reactants in photochemical reactions. For each solar case, we include photon fluxes from 0.5--2400 nm, binned in 1 nm increments. Total flux, once obtained, is scaled to Mars' orbit and SZA=60\degree.

We determined the dates of recent representative solar conditions by looking for periods when Ly $\alpha$ irradiance in the Lyman-alpha Model Solar Spectral Irradiance data set \cite{LISIRD} reached a peak, average, or trough. Because solar maximum and mean in recent decades have been historically quiet, we chose dates from the early 2000s to get a more representative photon flux for maximum and mean (solar minimum has not changed much). The dates we used were February 25, 2019 for solar minimum; February 7, 2004 for mean; and March 22, 2002 for maximum.

For the insolation flux data, we use SORCE/SOLSTICE at solar minimum and mean, and a mix of SORCE/SOLSTICE and TIMED/SEE at solar maximum. There is an additional complication for solar maximum: SORCE/SOLSTICE began a year after our solar maximum date, but includes the longer wavelengths we need, while TIMED/SEE began before our solar maximum date, but only includes fluxes at wavelengths shortwards of 190 nm. We patched together these two datasets, using SORCE/SOLSTICE for wavelengths 190-2000 nm from June 4, 2015 and TIMED/SEE for wavelengths 0.5-189.5 nm from March 22, 2002.

Figure \ref{inputs}a shows the fluxes only from 0.5 to 300 nm for simplicity; longwards of 350 nm, the profile does not vary over the solar cycle. The region shortward of 350 nm is also more important for photochemistry as the photodissociation and photoionization cross sections are largest there. We use the same cross sections as \citeA{Cangi2020}, with the addition of new photoionization and a few neutral photodissociation cross sections, the same used by \citeA{Vuitton2019}.

\subsection{Boundary conditions} \label{sect:bcs}

We use mostly the same boundary conditions as \citeA{Cangi2020}. The key addition is an additional non-thermal flux boundary condition at the top of the model for H, D, H\2, and HD, according to the functional form described by \cite{Gregory2023a}. Flux is zero at the top and bottom of the model for all ion species and any neutral species without a different boundary condition.

It is worth emphasizing that our flux boundary condition at the top of the model for atomic O is fixed at 1.2$\times 10^{8}$ \flux{}. This value has been used in past work \cite{Nair1994} to ensure that equilibrium models would produce H escape rates similar to that calculated from Mariner data. Results from MAVEN \cite{Dong2015, Leblanc2015, Brain2015, Lillis2017, Jakosky2018} indicate a present-day total O escape of closer to 3.8$\times 10^7$ \flux{}. We retain the use of 1.2$\times 10^{8}$ \flux{} in this work, but in future work will explore the effect of the lower MAVEN-derived loss rates on H and D escape. 

\section{Results} 

\subsection{What are the atmospheric densities of deuterated ions?}

\begin{figure}
    \includegraphics[width=\linewidth]{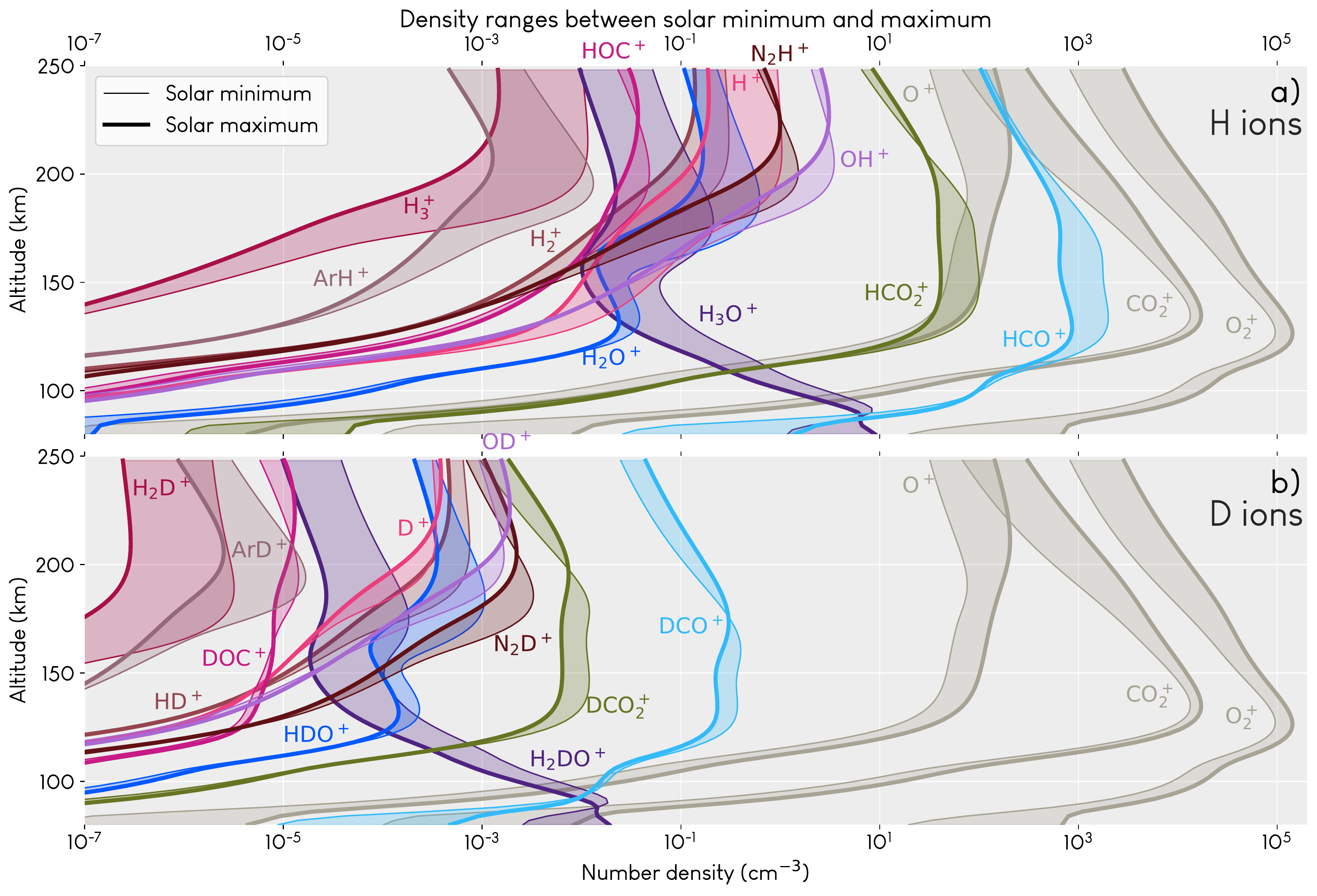}
    \caption{Densities of a) H-bearing ions and b) D-bearing (deuterated). Density ranges are bounded by their values at solar minimum (thin line) and solar maximum (thick line), with solar mean values falling within the shaded area. Gray lines show the primary ionospheric species for comparison. }
    \label{densities}
\end{figure}

The general distribution of the deuterated ionospheric species is similar to that of their H-analogues. Vertical profiles for select species containing H or D are shown in Figure \ref{densities}. Although they are calculated from surface to 250 km, the figure's lower boundary is placed at 80 km for legibility. The full image from surface to 250 km showing all species in the model appears in the Supporting Information (Figure S1). 

Primary peaks in the densities of deuterated ions occur between 150 and 200 km, with a minor peak near the top of the mesosphere, around 90-125 km. This structure does not hold for all species. H$_3$O$^+$ has its peak much lower down at about 90 km, which is in agreement with previous modeling \cite{Fox2015b, MolinaCuberos2002}. Unfortunately, comparisons with data are not feasible at this altitude because such data do not exist. Most ionic species, H- and D-bearing alike, also display a slight dip in density around 150 km, which is caused by a feature of the same shape in the electron temperature profile (see Figure \ref{inputs}a).

At solar maximum, greater insolation at short wavelengths enables more photoionization, increasing the abundances of primary species CO\2\p, O\2\p, and O\p which are produced directly from the parent neutrals. But for the lighter (and often more minor) ions containing H and D, chemistry and/or transport is a more important driver than photoionization. Temperature-driven changes in the parent neutral densities propagate through to their ions; for example, H$^+$ abundance at the top of the atmosphere decreases as the temperature goes up because H escape is diffusion-limited, whereas the same is not true for D abundance \cite{Cangi2020, Zahnle2008}. For other minor species that are not diffusion-limited, higher temperatures can also stimulate faster chemical reactions, slightly enhancing production and therefore density at higher temperatures.

\subsubsection{Comparisons with previous works}
Here, we compare our results to modeling results by \citeA{Fox2015b, Fox2021} and measurements by MAVEN NGIMS \cite{Benna2015, Fox2021}. In this work, we have parameterized our atmosphere in order to obtain an understanding of the mean-field behavior in time and space. We have not attempted to match the same  modeling input or the relevant atmospheric conditions of those studies. Our models differ substantially from those by \citeA{Fox2021, Fox2015b} in temperature structure, boundary conditions (especially for ions at the upper boundary), vertical extent, use of photochemical equilibrium, background atmosphere, SZA, included species, mean Mars-Sun distance, assumed eddy diffusion profile, and included processes (we do not model electron impact ionization or dissociation). Because of these differences, we provide these comparisons primarily for the reader's orientation. 

\begin{figure}
	\includegraphics[width=\linewidth]{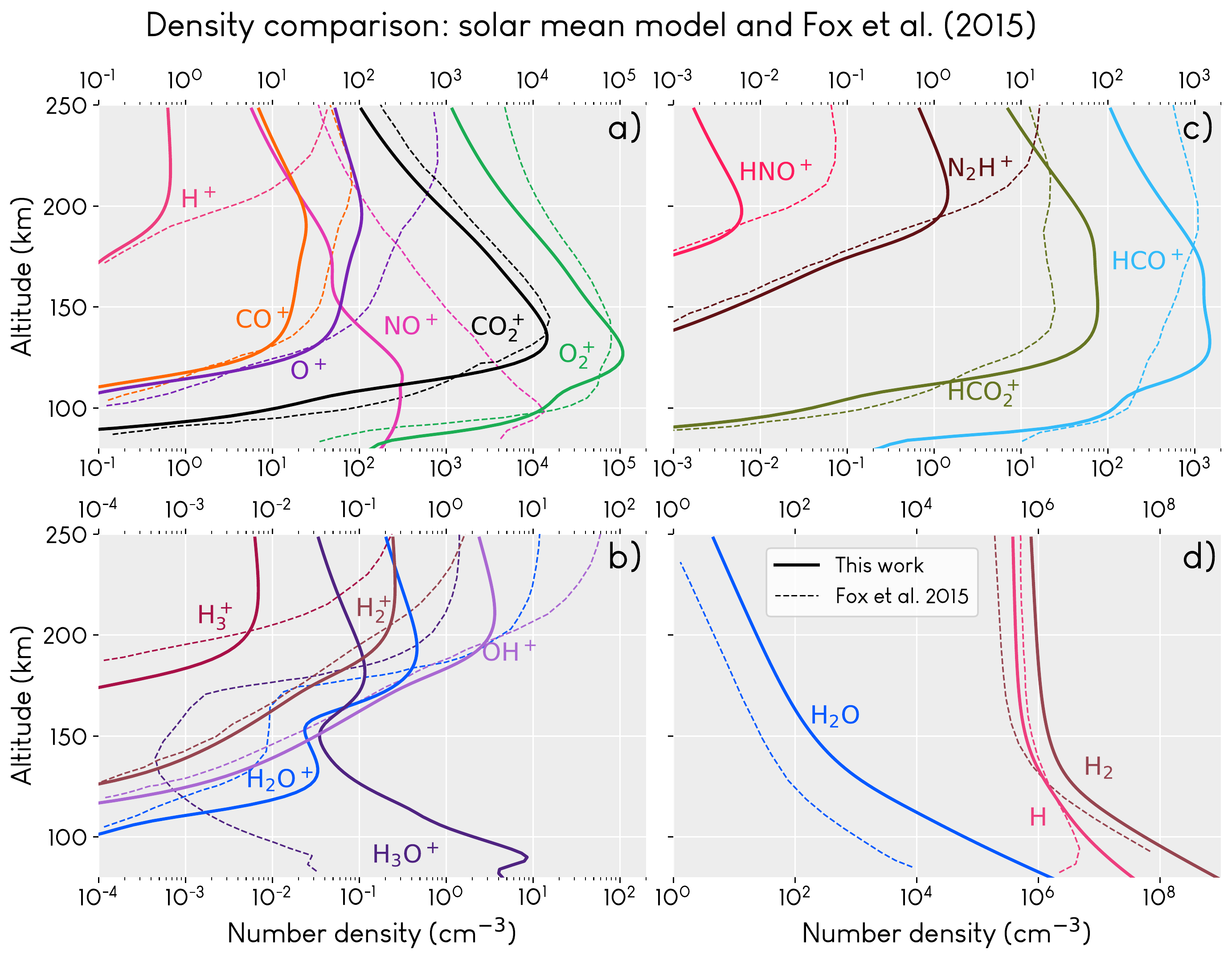}
	\caption{Ion and neutral densities computed by our model and compared with those computed by \citeA{Fox2015b}. Species are divided amongst the four panels for legibility and compared with Figure 3 in \citeA{Fox2015b}. Some minor species are omitted for clarity.}
	\label{fig:compfox}
\end{figure}

\textbf{\citeA{Fox2015b}}. For the major ions such as O\p, CO\2\p, and O\2\p, our density profiles are generally consistent with those modeled by \citeA{Fox2015b}, as shown in Figure \ref{fig:compfox}. They are also broadly similar for many of the minor ions, although in general, our profiles tend to show lower densities near 250 km by 1-2 orders of magnitude. There is a significant difference between our NO$^+$ profiles, which likely relates to differences between this work and that of \citeA{Fox2015b} in the nitrogen ion chemistry and included species. It should be noted that many of the ions for which we show a significantly different profile are quite minor, with populations never exceeding 100 cm$^{-3}$, so the absolute differences as a percent of the total atmosphere are tiny, well within the absolute tolerance. \citeA{Fox2015b} make the point that their model calculates neutral H\2O produced only by ion-neutral reactions due to their choice of boundary conditions, whereas ours includes production by photodissociation and lower atmospheric neutral chemistry; it is then perhaps not surprising that our results include more water and different distributions of water-group ions and H-bearing species than theirs (see Figure \ref{fig:compfox}d).

In Figure S3, we also compare our results to \citeA{Fox2021}, which uses a similar model to \citeA{Fox2015b} and includes recent data from NGIMS for CO\2\p, O\2\p and O\p. Compared to that paper, our results are more dissimilar.

\begin{figure}
	\includegraphics[width=\linewidth]{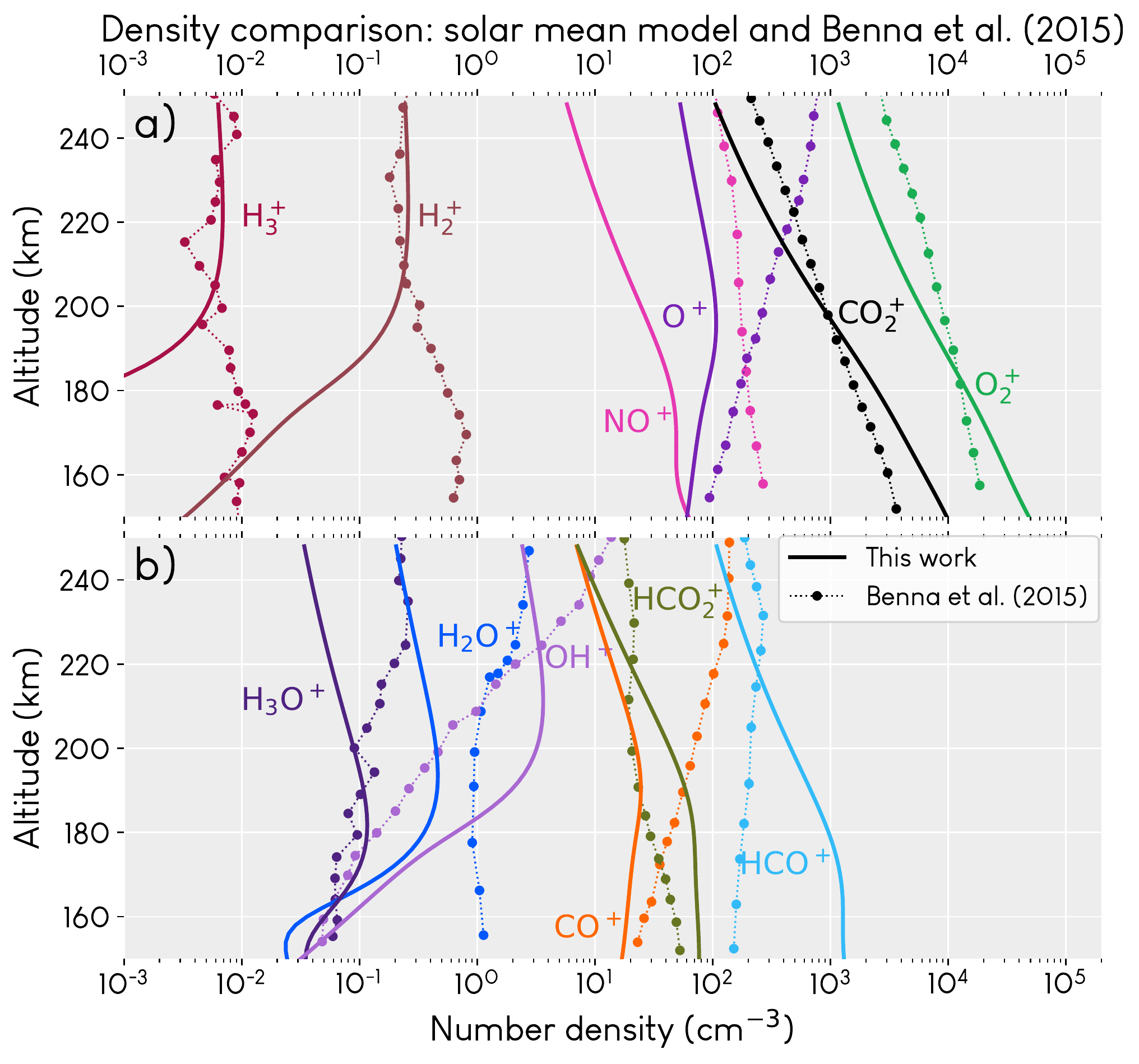}
	\caption{Ion and neutral densities computed by our model and compared with those computed by \citeA{Benna2015}. Species are divided amongst the two panels for legibility. HNO\p differs significantly from data and has been omitted; the measurements are known to be unreliable due to spacecraft potential.}
	\label{fig:compbenna}
\end{figure}

\textbf{\citeA{Benna2015}, using MAVEN NGIMS}. Figure \ref{fig:compbenna} shows that our results are in reasonably good agreement with the initial NGIMS measurements at Mars \cite{Benna2015}, which occurred long enough into the mission that solar mean conditions would have prevailed. There continues to be a divergence between model and data for O\p in the upper atmosphere and an underprediction of NO\p, but considering we are using a 1D model that does not account for local and short-term variations and we have not made any model changes to match data, we find the output acceptable. Significant differences in minor H-bearing ions have a variety of explanations here. NGIMS has background noise typically at 0.01 cm$^{-3}$ and below, and a random uncertainty of 50\% at 0.1 cm$^{-3}$ \cite{Benna2015}, which may explain the divergence in H\3\p and H\2\p; this difference could also result from the larger amount of the principal H\2O\p ion present in the data. H\2\p and H\3\p are also measured in the lowest mass channels that the instrument can access, where the instrument is less sensitive due to a correlation of sensitivity with cross section; there is also a possibility of H outgassing from the instrument fillament \cite{Mahaffy2015}.

\subsection{Are the dominant production mechanisms of hot H and D analogous or dissimilar?}

\begin{figure}
\includegraphics[width=\linewidth]{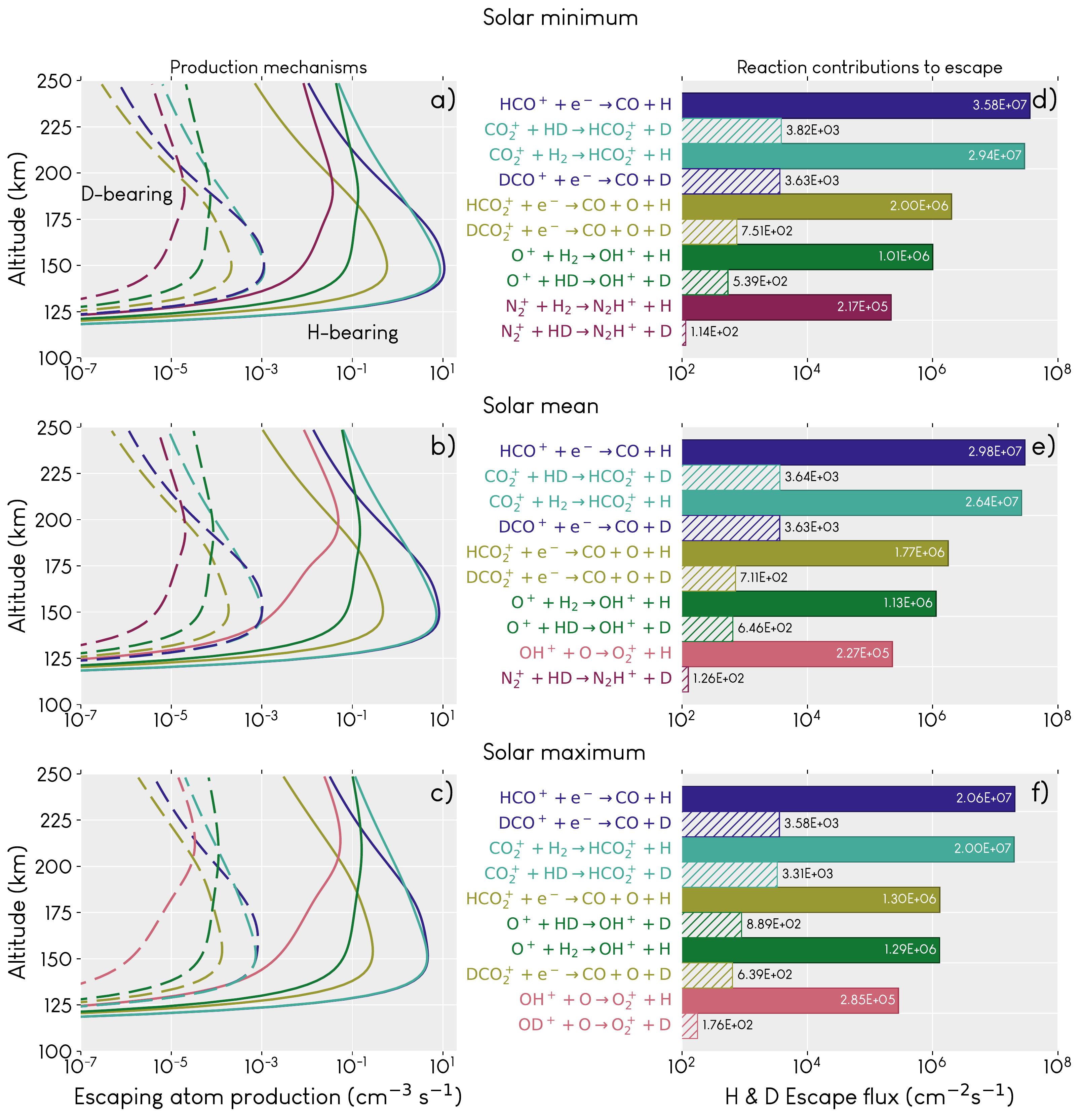}
\caption{Volume production rates of escaping atoms (panels a, b, c) and integrated escape flux of the produced atomic H or D (d, e, f) for the dominant five chemical pathways producing hot H (solid lines/solid bars) and hot D (dashed lines/striped bars).}
\label{mech}
\end{figure}
Figure \ref{mech} shows the production mechanisms for hot H and D, which are mostly similar. The hot H production mechanisms are also calculated independently by \citeA[this issue]{Gregory2023b}. 

The most important reaction driving the production of hot D (H) below 200 km in solar mean and maximum is DCO\p (HCO\p) dissociative recombination (DR), with CO\2\p + HD (H\2) a close second. HCO$^+$ DR dominates for hot H under all solar conditions, but for hot D, CO\2\p + HD marginally dominates over DCO$^+$ DR during solar minimum at certain altitudes, making it the dominant source of escaping hot D at solar minimum. This is because the density of HD relative to DCO$^+$ is larger than H\2 relative to HCO\p. The rates of production from these two processes for hot D are very close; minor changes in conditions, including normal fluctuations in the real atmosphere, could likely change this relationship. Above 200 km, CO\2\p + H\2 dominates for hot H production, but high-altitude hot D comes mostly from O\p + HD.

DCO\2\p (HCO\2\p) DR is the third most important reaction during quiet solar conditions, but it is eclipsed by O+ + HD (H\2) during solar maximum. Under quieter solar conditions, the fifth place position is seized by N\2\p + HD (H\2). But as the thermosphere warms, OD (OH) + O claims the fifth place, first for the H species and then for the D species. This appears to be because the dominant reaction involving OH\p and OD\p is the reaction O\p + H\2 (HD) $\rightarrow$ OH\p (OD\p) + H. This reaction also has a rate coefficient that is independent of temperature, whereas N\2\p + HD (H\2) has a rate coefficient which decreases with temperature.

\subsection{What is the magnitude of non-thermal escape of D, and under which conditions does it dominate thermal escape?}

\begin{figure}
    \includegraphics[width=1\linewidth]{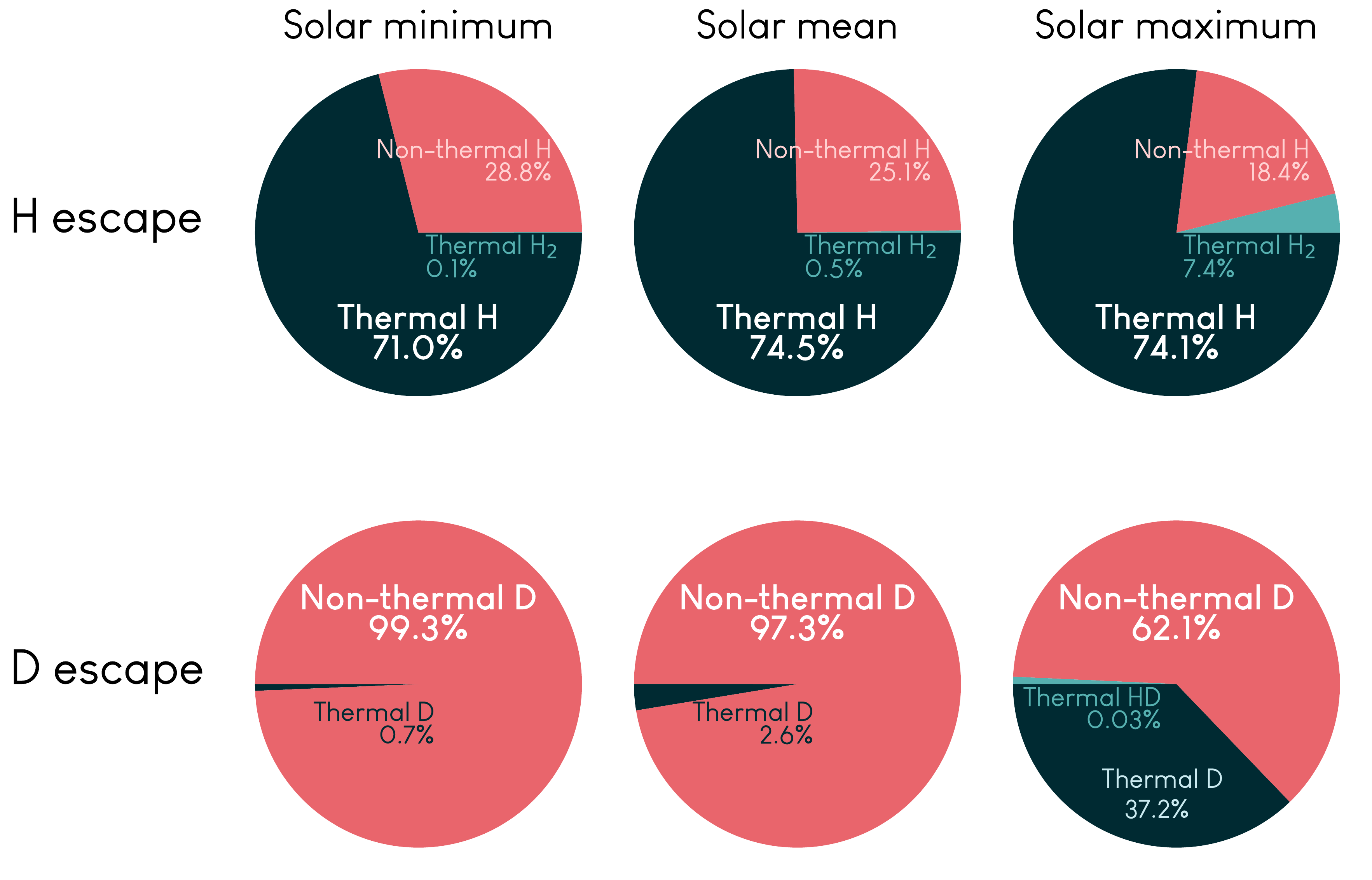}
    \caption{Relative escape contributions for H and D. As expected based on the literature, thermal escape dominates for H during all solar conditions, but  non-thermal escape dominates D escape, even at solar maximum. Although we do model non-thermal escape of H\2 and HD, their contributions are completely negligible (see Table \ref{tab:fluxes}).}
    \label{fig:escbal}
\end{figure}

Figure \ref{fig:escbal} shows the relative contributions of thermal and non-thermal escape of atomic H and D and thermal escape of the molecular species; the associated escape fluxes to space are given in Table \ref{tab:fluxes}. The density profiles of the neutral species, from which the escape is sourced, appear in Figure S2; an upcoming publication will focus on variations in these neutral species and their D/H ratios. As in previous work, \cite<e.g.>{Kras2002}, thermal escape is the dominant loss process for atomic H, with non-thermal escape of H making up a gradually reducing share as solar activity increases. The picture looks very different for D, for which 62-99.3\% of escape is non-thermal depending on solar conditions. Note that, as shown in Table \ref{tab:fluxes}, the total escape of H and D adds to 2.4$\times 10^{8}$ \flux{} under all solar conditions due to the boundary condition on O escape. H and O escape are linearly related when the atmosphere is in equilibrium: the sum $\mathrm{\phi_H+\phi_D}$ will naturally evolve to equal twice the O escape flux, since H\2O and HDO are the primary source of H and D in the model, and contain two H or D for every O. This relation does not necessarily hold over shorter timescales, when short-term perturbations can temporarily break this stoichiometric equilibrium.

Previous work has predicted that thermal escape of D should actually dominate at solar maximum \cite{Kras2002} and that non-thermal escape of D in the form of larger molecules such as HD, OD, and HDO could be up to 15\% \cite{Gacesa2018}, whereas our results show that non-thermal escape of HD is so negligible as to not appear at all in Figure \ref{fig:escbal}. Besides the fact that we do not account for excited rotational states of HD, the discrepancy also likely arises from our chosen methods. Our non-thermal escape probability curve is valid for hot H atoms with 5 eV of energy, and we do not account at all for branching to excited internal states of the other products; we assume that all atomic H and D produced by exothermic reactions are produced ``hot''. In reality, not all exothermic heat is dumped directly into the lone atoms all the time. With proper accounting for these intricate branching ratios, our calculated total of non-thermally escaping atomic D would likely decrease. We also do not calculate non-thermal OD escape, as escape probabilities for a molecule of this mass are not available.

\begin{table}[] 
\resizebox{\textwidth}{!}{%
\begin{tabular}{llllllllllll} \toprule
 & 
  \multicolumn{4}{c}{Thermal escape (\flux{})} &
  \multicolumn{4}{c}{Non-thermal escape (\flux{})} &
  \multicolumn{3}{c}{Total escape (\flux{})} \\ \cmidrule(r){2-12}
 &
  \multicolumn{1}{c}{H} &
  \multicolumn{1}{c}{D} &
  \multicolumn{1}{c}{H$_2$} &
  \multicolumn{1}{c}{HD} &
  \multicolumn{1}{c}{H} &
  \multicolumn{1}{c}{D} &
  \multicolumn{1}{c}{H$_2$} &
  \multicolumn{1}{c}{HD} &
  \multicolumn{1}{c}{H} &
  \multicolumn{1}{c}{D} &
  \multicolumn{1}{c}{H + D} \\ \cmidrule(r){2-12}
Solar minimum & 1.70\E{8}  & 62  & 1.66\E{5} & 0.1 & 6.92\E{7} & 9225 & 11871 & 7 & 2.39991\E{8} & 9295  & 2.4\E{8} \\
Solar mean    & 1.79\E{8} & 243  & 5.43\E{5} & 0.9 & 6.02\E{7}  & 9154 & 9491  & 7 & 2.39991\E{8} & 9405  & 2.4\E{8} \\
Solar maximum & 1.78\E{8} & 5415 & 8.92\E{6} & 104  & 4.43\E{7}  & 9041 & 5473  & 4 & 2.39985\E{8} & 14564 & 2.4\E{8} \\ \bottomrule
\end{tabular}
} 
\caption{Amount of thermal and non-thermal escape of atomic and molecular H and D species for the three solar conditions. The total escape amounts to 2.4$\times10^8$ \flux{} because in the equilibrium atmosphere, the ratio $\phi_H/\phi_O$ approaches 2, as O escape is fixed at 1.2 $\times 10^{8}$ \flux{} (see Section \ref{sect:bcs}). Escaping atoms and molecules are sourced from the neutral species; densities for the associated species are shown in Figure S2.} \label{tab:fluxes} 
\end{table}

\section{Discussion}

\begin{figure}
    \includegraphics[width=0.6\linewidth]{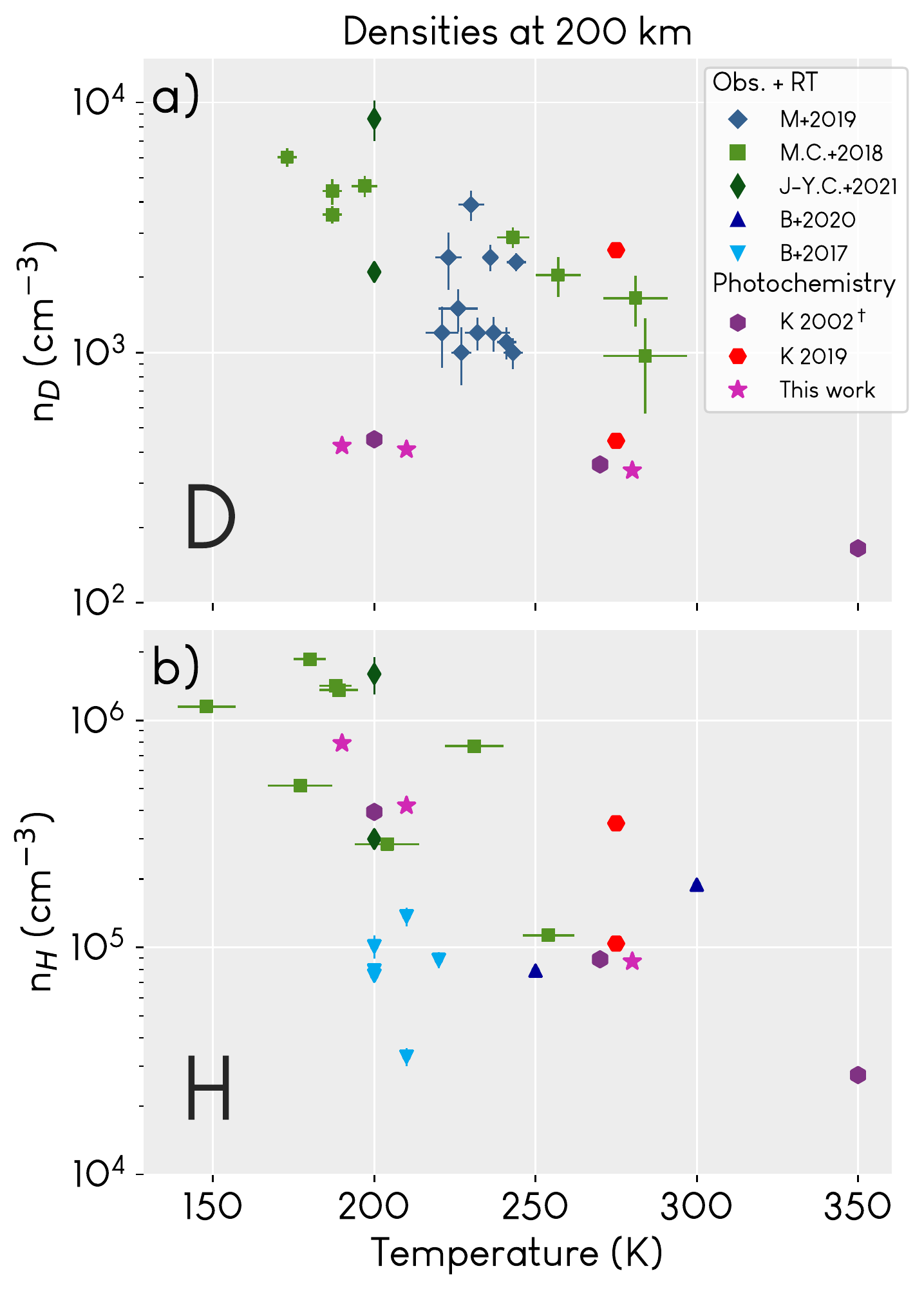}
    \caption{D and H densities at 200 km ($^\dagger$ 250 km) from multiple studies. Data represent multiple solar zenith angles, seasons, hemispheres, etc. M+2019: \citeA{Mayyasi2019}. M.C.+2018: \citeA{Chaffin2018}. J-Y.C.+2021: \citeA{Chaufray2021b}. B+2020, 2017: \citeA{Bhattacharyya2020, Bhattacharyya2017}. K 2002, 2019: \citeA{Kras2002, Kras2019}. Entries under ``Obs. + RT'' used brightness observations from either HST \cite{Bhattacharyya2017} or MAVEN IUVS (all others) with radiative transfer modeling for density retrievals. For these studies, invisible density error bars indicate uncertainty smaller than the marker size. Temperature error bars indicate that temperature was retrieved from spacecraft data, while missing temperature error bars mean it was a model parameter or output. Uncertainties for photochemistry studies are not calculated. Photochemical modeling typically reports an order of magnitude less D than other methods, which may be due to observation biases toward times of brighter D emission. There is no similar discrepancy in H densities.}
    \label{fig:multistudy_densities}
\end{figure}

Figure \ref{fig:multistudy_densities} places our D and H densities in context with other studies. We have only consolidated reported densities; we make no attempt to filter by observation geometries. Nevertheless, there appears to be an inverse relationship of densities and temperature for both species. We can also see that photochemical models (red/purple/pink points) produce D densities that are an order of magnitude smaller than densities retrieved using observations and radiative transfer modeling; the same discrepancy does not occur for the H densities. Deuterium Lyman $\alpha$ is difficult to separate from hydrogen Lyman $\alpha$; the D density discrepancy may potentially be explained by a systematic bias toward anomalously bright D emissions. One exception is the density of D at $\sim$2500 and T=$275$K in the work by \citeA{Kras2019}; this point represents a model run with a high amount of water in the thermosphere, whereas all the other photochemical results have a comparatively lower water abundance. This comparison demonstrates that our model output is in reasonable agreement with other works. 
 
As mentioned previously, we do not include cloud or dust microphysics, although these processes do have an important effect on the water cycle. These effects are explored in two recent papers using the Laboratoire de Météorologie Dynamique Planetary Climate Model (LMD-PCM) to study the creation of water ice clouds and their role in controlling the D/H ratio \cite{Vals2022, Rossi2022}.

\subsection{Can inclusion of non-thermal escape in the model yield an estimation of water loss similar to the amount calculated in geomorphological studies?}

By considering both thermal and non-thermal escape, we can now compute the D/H fractionation factor, which represents the relative efficiency of D and H escape. It is defined as: 
\begin{equation}
    \mathrm{f = \frac{\phi_{D}/\phi_{H}}{[HDO]_s/2[H_2O]_s}}
\end{equation}
Where $\phi_X = \phi_{X,t} + \phi_{X, n}$ is the rate at which species X (D or H) escapes from the top of the atmosphere due to both thermal ($t$) and non-thermal ($n$) processes. The denominator represents the D/H ratio in water measured at the surface ($s$), which is a proxy for the D/H ratio in the larger exchangeable reservoir. 

The fractionation factor is important not only because it tells us how efficient loss of D is compared to loss of H, but also because it is useful for calculating the integrated water loss from a planet. Long-term enrichment of the heavy isotope (D) due to differential escape of D and H can be modeled using Rayleigh fractionation \cite{ChamberlainHunten1987, Yung1998}:

\begin{equation} \label{eqn:rf}
    \mathrm{\frac{(D/H)_{now}}{(D/H)_{past}} =\left(\frac{[H]_{past}}{[H]_{now}}\right)^{1-f} }
\end{equation}

Equation \ref{eqn:rf} is used to calculate water loss from Mars. The D/H ratio on the left hand side represents the ratio measured in water in the exchangeable reservoir (the seasonal polar caps, near-surface ices, and atmospheric water vapor), and the ratio H\2O$\mathrm{_{past}}$/H\2O$\mathrm{_{now}}$ can be substituted in on the righthand side and rearranged, obtaining \cite{Cangi2020} (where W is water):

\begin{equation} \label{eqn:rf-water}
    \mathrm{W_{lost} = W_{now} \left( \left( \frac{(D/H)_{now}}{(D/H)_{past}} \right)^{1/(1-f)} - 1\right)}
\end{equation}
Implicit in these equations is the assumption that [H] $\gg$ [D], so that the past and present abundances of H\2O are reasonable representations of the entire water budget. In the present day, the ratio of D/H is well constrained by many observational studies to be approximately 4-6 $\times$ standard mean ocean water (SMOW) \cite[and references therein]{Encrenaz2018, Villanueva2015}. Current research also has identified a likely present-day exchangeable reservoir water budget of 20-30 m GEL \cite[and references therein]{Lasue2013}. By obtaining a reliable value for $f$, we can combine all these values to calculate the inventory of water on ancient Mars.

\begin{figure}
    \includegraphics[width=\linewidth]{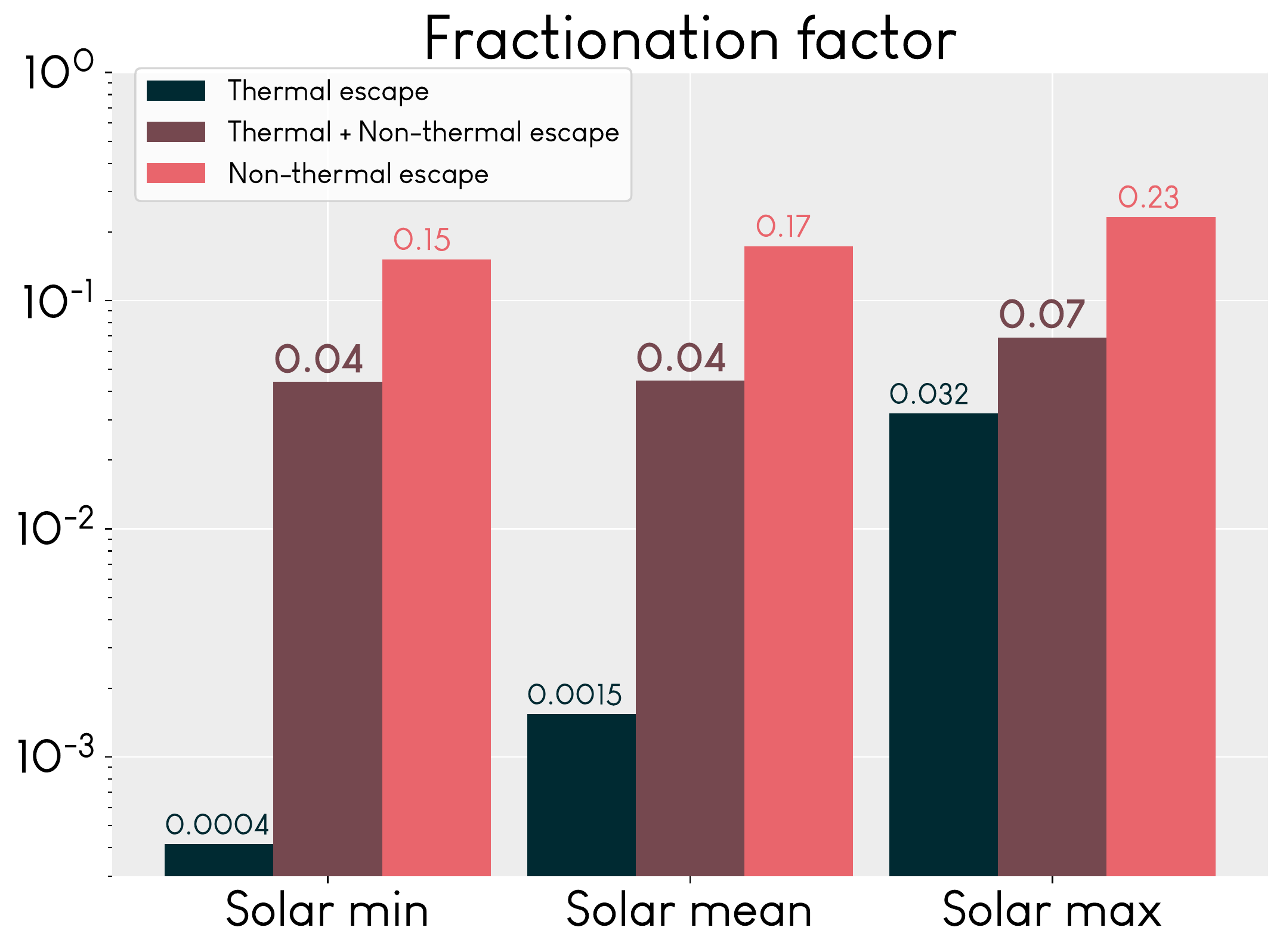}
    \caption{The fractionation factor $f$ for three different modes of escape. Increased solar activity leads to amoderate increase in total $f$, and inclusion of non-thermal escape increases total $f$ by 1-2 orders of magnitude during solar minimum and mean. Fractionation represents the escape efficiency of D compared to H, meaning that $f=0.04$ represents a 4\% escape efficiency of D. Non-thermal escape is an effective escape method for D under all solar conditions. }
    \label{fig:ff}
\end{figure}

\citeA{Cangi2020} suggested that the difference between the mean atmospheric $f_t$ (considering only thermal escape) and $f_{tn}$ (considering both thermal and non-thermal escape) was several orders of magnitude. Because they did not directly model non-thermal escape, they arrived at this conclusion by incorporating the non-thermal escape velocity given by \citeA{Kras1998} into their model. We are now in a position to compare with those estimates; our calculations of the fractionation factor are shown in Figure \ref{fig:ff}. \citeA{Cangi2020} calculated $f=0.06$ for their standard atmosphere, based on their modeled thermal escape and estimated non-thermal escape. We calculate a total escape fractionation of $f=0.04$ for our solar mean atmosphere, which has the same insolation and similar temperatures, and is not far off from their 0.06. Our results are consistent with their thermal escape $f=0.002$ for the standard atmosphere (roughly equivalent to our solar mean atmosphere). Our results show that while overall D escape at Mars is around 4-7\% as efficient as H escape, non-thermal D escape is much more efficient, between 15-23\% that of H. 

\begin{figure}
    \includegraphics[width=\linewidth]{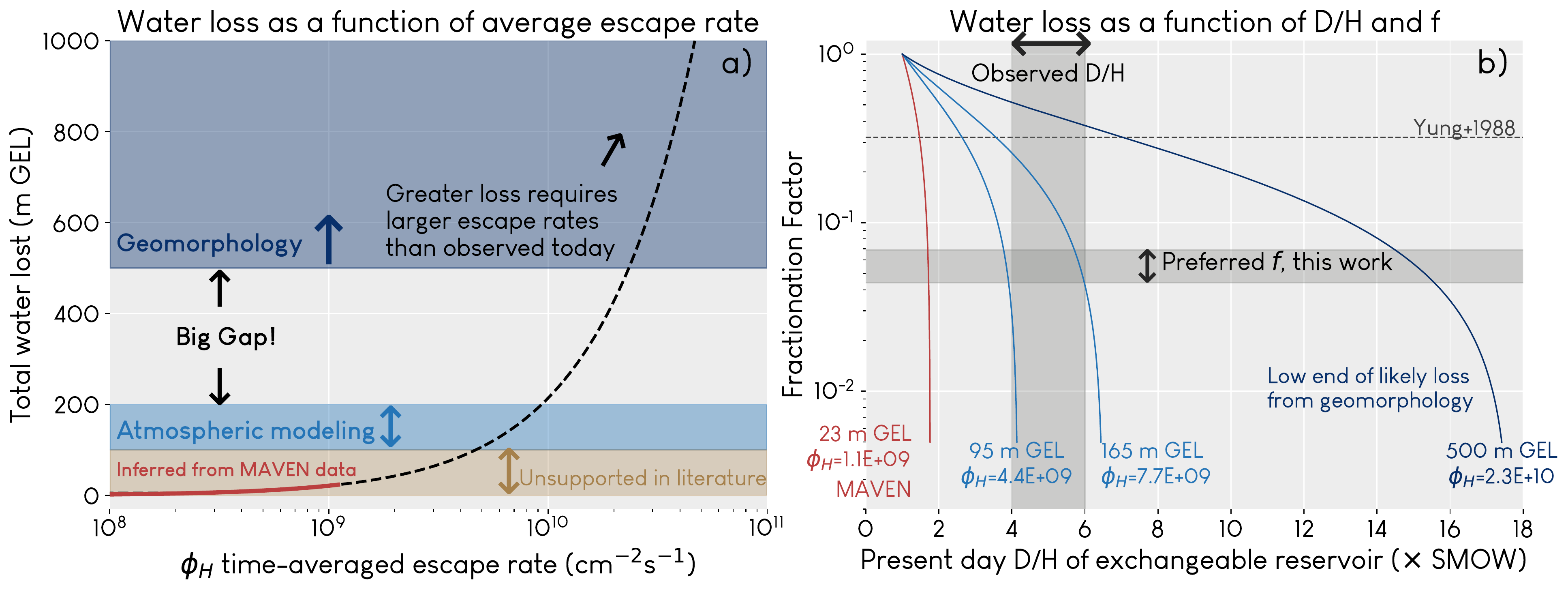}
    \caption{a): Possible water loss as a function of long-term average H escape rate $\phi_H$, $\mathrm{W_{lost}=\bar{\phi_H}}t$, where $t=4.5$ billion years. A significant gap separates the amount of water loss inferred from atmospheric modeling and geomorphological studies. Additionally, escape rates determined from MAVEN data enable very small amounts of water loss that are not consistent with the geological evidence. b): Water loss lines represent solutions to equation \ref{eqn:rf-water}, assuming 30 m GEL in the present-day exchangeable reservoir. The regions matching the best values of D/H and $f$ are shaded in gray, with the overlapped rectangle representing our best estimate of the present-day atmosphere. (The fractionation factor calculated by \citeA{Yung1988} is shown for reference, though it is high due to the highly uncertain exospheric temperatures then used.)}
    \label{fig:thedisagreement}
\end{figure}

Our results yield integrated water loss of 147--158 m GEL (present day exchangeable reservoir = 30 m GEL, $f=0.04$--$0.07$, D/H=$5.5\times$SMOW). This total loss still does not agree with the geological estimates of 500+ m GEL \cite{Lasue2013}. The discrepancy is summarized in Figure \ref{fig:thedisagreement}. Figure \ref{fig:thedisagreement}a shows the gap between the amount of water loss calculated by atmospheric models \cite{Yung1988, Kass1999, Kras2000, Kras2002, Cangi2020} and that inferred from geomophological observations \cite[and references therein]{Lasue2013}. The time-averaged H escape rate curve suggests that the rates observed today \cite{Jakosky2018} are unlikely to be near the average, and that escape was likely higher in the distant past, enabling greater water loss. Plausible explanations could include periods of hydrodynamic escape, a more EUV-active young sun driving greater photochemistry, extreme obliquities \cite{Wordsworth2016, Laskar2004}, or other as of yet unknown dynamics. 

It is also possible that some water may have been sequestered into the surface. Recent work by \citeA{Scheller2021} suggests that this amount may have accounted for between 30-99\% of all missing water. More smaller-scale models and many observations will be needed to constrain this large range further. Hydrated minerals may contain 130-260 m GEL equivalent water \citeA{Wernicke2021}, but the time of emplacement and any fractionation of the process is unclear. In general, due to the chaotic evolution of obliquity \cite{Laskar2004} over Mars' history, it is extremely difficult to qualitatively describe escape rates in the past. Although it is difficult to extrapolate much from the present-day rates, high loss of water via escape to space is not ruled out. 

Figure \ref{fig:thedisagreement}b also helps demonstrate when it is important to know the value of $f$ rather precisely. Discriminating between $f=0.04$ or $f=0.07$ is not particularly important: below $f=0.1$, water loss curves are relatively vertical, meaning that a change in $f$ does not equate to a significant change in water loss, but this is less true the closer $f$ gets to 1. (For another view, see Figure S4 for water loss as a function of $f$ for a single D/H ratio.)

Considered together, these insights tell us that non-thermal escape processes for D are important to model in order to accurately understand how D escapes from Mars. This conclusion may not hold for other planets, moons, or exoplanets; on bodies which are colder, larger, or otherwise less conducive to thermal escape, non-thermal escape may have a greater role to play. 

\subsection{Other non-thermal processes}

As mentioned earlier, we do not model interactions with the solar wind. \citeA{Lewkow2014} have calculated that penetrating solar wind protons can produce $4.18 \times 10^4$ \flux{} energetic neutral H atoms (ENAs) which escape. This is a negligible amount compared to our total non-thermal H escape, which is $6.5\times 10^7$ \flux{}, so we do not expect it to change our results. Without a value for D ENAs, we cannot make an immediate comparison, but we can calculate what fraction of our total H column the H ENAs make up. For our solar mean model (Figure S1), the fraction is 8$\times 10^{-11}$. If we multiply this fraction by our D column, we obtain an additional 25 \flux{} non-thermally escaping D atoms. Solar wind penetration should thus be negligible for the non-thermal escape we calculate at the altitudes we model, but an important interaction to consider during times of extreme solar activity or at higher altitudes in the exosphere. A future study on D ENA production would also help refine this contribution's value.

We do not account for the collision of hydrogen atoms or molecules with hot oxygen, which is another significant source of hot atoms in the martian atmosphere. Assuming an exospheric temperature of 240 K, \citeA{Gacesa2012} calculated that $1.9 \times 10^5$ \flux{} H\2 molecules escape as a result of collision with hot oxygen, which is larger than our non-thermal H\2 flux by 1-2 orders of magnitude depending on solar conditions (see Table \ref{tab:fluxes}). They also estimate that 74 HD molecules \flux{} escape via this mechanism. This would bring our total HD escape to approximately 100 \flux{}, an order of magnitude larger than our current result. Other species may also play a role; \citeA{Gacesa2017} calculate that the total non-thermal escape of OH is $1.07\times10^{23}$ s$^{-1}$, i.e. $7.4 \times 10^5$ \flux{}. Even added together, these numbers are all still orders of magnitude smaller than the non-thermal atomic escape fluxes, and will not significantly affect our results. If we included them, the net effect would be to boost H escape, decreasing the fractionation factor and total water loss.

Energization of atomic H and D by collision with hot oxygen may be more significant. \citeA{Shematovich2013} estimates, for specific density profiles and temperatures, a total possible escape flux of hot H produced this way to be $6 \times 10^6$ \flux{} at low solar activity. This is 9\% of our non-thermal H escape (see Table \ref{tab:fluxes}). Our non-thermal D escape is 3 orders of magnitude lower than the H escape. If we  apply this scaling relation to hot O collisions with D, we can expect that this pathway might produce D escape on the order of $10^3$, which is the same order as our calculated non-thermal escape fluxes. If we instead generously assumed an extra $9 \times 10^3$ hot D atoms produced from collision with hot O, doubling our total, we would expect to also double our fractionation factor, to $f=0.08-0.14$.

\subsection{Future opportunities and directions}

There are several things that could enhance our model. The first likely avenue worthy of exploration would be to perform a similar study, but with a more physically-motivated parameterization of atomic O escape. Fixing the O escape at $1.2\times 10^{8}$ \flux{} was sufficient for the scope of this work; our results represent long-term equilibrium, when it is possible to adopt reasonable means for parameters like O escape. Adding a dynamically evolving escape flux boundary condition for atomic O would enable a more comprehensive understanding of shorter-term variations in H and D escape rates, such as a result of regular seasonal cycles. This would better capture the interplay between the hydrogen species and CO\2, the main component of the atmosphere and a significant source of O. This would also present an opportunity to include processes more important to O loss, such as ion pickup, ion/polar outflow, and sputtering. We do not include these as we focus on H and D loss, which are dominated by other processes.

We have also been forced to make some unavoidable assumptions about the basic chemistry, owing to a lack of laboratory data. While we have made a best attempt to use existing reaction rate coefficient data from several different papers and databases, a comprehensive catalogue of rate coefficients, branching ratios, and cross sections for deuterated reactions is not available in the literature at this time. Most especially, future photochemical models would benefit from accurate photoabsorption cross sections for deuterated neutrals other than HDO (including OD and HD in particular), and measured reaction rate coefficients for as many of the deuterated reactions with estimated rates in Table \ref{tab:DREACTIONS} as possible. While not all reactions will significantly affect the chemistry, certain rates that dominate production or loss of a species can have strong effects, affecting densities up to a few orders of magnitude (see, for example, \citeA{Fox2017}).

Photochemical modeling often entails excluding some important processes that are better captured in higher-dimensional models. Bluejay is the first photochemical model to couple the ion and neutral atmospheres from the upper atmosphere down to the surface, but there is still an opportunity for future work to give more attention to surface-atmosphere interactions. Our inclusion of surface-atmosphere interactions is primarily relegated to surface density boundary conditions for certain species. A more detailed parameterization of processes such as volcanic outgassing, major seasonal changes in the polar caps, water adsorption and desorption on dust grains and dust lifting, deposition of volatiles, and the role of non-volatiles such as perchlorates, salts, and other non-water ices could yield new insights into the planetary climate system as a whole. 

Our results also have implications for the detectability of deuterated ions by present and future Mars missions. Using MAVEN's NGIMS instrument, the deuterated ions that we model typically occupy the same mass/charge ratio bin as a more prevalent H-bearing species. For example, D\p occupies the same bin as H\2\p, but the latter is far more abundant. The deuterated species in our model which do not overlap with an H-bearing species are H\2D\p (mass bin 4), HD\2\p (5), H\2DO\p (20), HDO\2\p (35), and ArD\p (42). However, several of these species are expected to be very rarefied and thus difficult to detect, and others may overlap with species we do not model that do exist on Mars, such as helium in mass bin 4.  These degeneracies make obtaining deuterated ion densities challenging; doing so will require inventive methods applied to existing data or new methods with new instruments. 

\section{Conclusions}

We have used a 1D photochemical model, bluejay, that fully couples ions and neutrals from surface to space to study production of hot D from planetary ionospheric processes. We show that the deuterated ionosphere behaves relatively similar to the H-bearing ionosphere. This result is somewhat expected, as measurements of rate coefficients for deuterated reactions are much less available than the H-bearing counterpart reaction rate coefficients. 

For the first time, we have self-consistently quantified, in raw flux and in percent of total escape, the thermal and non-thermal escape fluxes of H, D, H\2, and HD sourced from planetary ionospheric reactions under different solar conditions. We also show the first identification of the dominant chemical reactions which produce hot D. Our results confirm earlier suggestions that non-thermal escape dominates D escape at Mars, although our results have shown that this is true throughout the solar cycle rather than just during quiet solar conditions. 

We also confirm an earlier prediction \cite{Cangi2020} that including non-thermal escape when calculating the D/H fractionation factor will result in a fractionation factor several orders of magnitude higher than if it is neglected. However, the resulting fractionation factor is 0.04--0.07, meaning that D escape is only about 4-7\% as efficient as H escape. If the fractionation has consistently been this small, and we also assume that $\phi_H$ has been similar to the value today through time, it is difficult to ascribe the large amount of water loss that we see indicated in the rock record to atmospheric escape alone. On the other hand, the dust storm season on Mars, as well as normal seasonal variations between perihelion and aphelion, are characterized by spatially and temporally localized enhancements of the D/H ratio, water abundance, and H escape \cite[and references therein]{Villanueva2022, Daerden2022, Fedorova2021, Chaffin2021, Holmes2021, Fedorova2020, Stone2020, Aoki2019, Vandaele2019,Heavens2018, Chaffin2017}. It is not yet clear if enhanced D escape or a heightened fractionation factor also occur along with these seasonal changes, although it seems likely \cite{Alday2021}; if they do, then the assumption of a constant fractionation factor over time cannot hold, and we will have to introduce some additional nuance to our use of Rayleigh fractionation to estimate water loss.

Ongoing improvements in modeling, especially coupling between 1D and 3D models, as well as continual advancements in instrumentation for planetary missions will be necessary to continue putting together the puzzle of water on Mars throughout history.

\section{Open Research Statement}

The photochemical model, bluejay, used for this work is written for and compatible with Julia 1.8.5 \cite{Julia}. The model itself, in version 1.0 as used in this work, is available at Zenodo \cite{Cangi2022}. 

A typical use-case of the model is to modify simulation parameters within \texttt{PARAMETERS.jl} and to then call 
\texttt{julia converge\_new\_file.jl} at the command line.


\acknowledgments

This work was supported by three separate funding sources. First, this work was supported by Mars Data Analysis Program grant \#NNX14AM20G. Second, this material is based upon work supported by the National Science Foundation Graduate Research Fellowship Program under Grant No. DGE 1650115. Any opinions, findings, and conclusions or recommendations expressed in this material are those of the author(s) and do not necessarily reflect the views of the National Science Foundation. Third, this work was supported by NASA’s FINESST Program (Grant \#80NSSC22K1326).


\nocite{*}
\bibliography{Cangi2023}

\end{document}